\documentclass[11pt]{article}

\usepackage[utf8]{inputenc}
\usepackage{times, comment}
\usepackage{setspace, authblk, subcaption, textcomp}
\usepackage{amsmath,enumitem, amsfonts}
\usepackage{mathtools,bm}
\usepackage{amsthm,color, booktabs, multirow}
\usepackage{natbib}
\usepackage{float,placeins}
\usepackage{chngcntr} 

\usepackage[margin=1in]{geometry}
\DeclareMathOperator{\bX}{\mathbf{X}}
\DeclareMathOperator{\bZ}{\mathbf{Z}}
\DeclareMathOperator{\bbeta}{\bm{\beta}}
\DeclareMathOperator{\bW}{\mathbf{W}}
\DeclareMathOperator{\bU}{\mathbf{U}}

\DeclarePairedDelimiter\ceil{\lceil}{\rceil}
\DeclarePairedDelimiter\floor{\lfloor}{\rfloor}
\DeclareMathOperator*{\argmin}{argmin} 

\title{Screening methods for linear errors-in-variables models \\ in high dimensions}
\newtheorem{theorem}{Theorem}
\newtheorem{lemma}{Lemma}
\author[1,3]{Linh H. Nghiem}
\author[1]{Francis K.C. Hui}
\author[2,3]{Samuel M{\"u}ller}
\author[1]{A.H. Welsh}
\affil[1]{Research School of Finance, Actuarial Studies and Statistics, Australian National University, Australia}
\affil[2]{Department of Mathematics and Statistics, Macquarie University, Australia}
\affil[3]{School of Mathematics and Statistics, University of Sydney, Australia}

\date{}                     
\begin{document}
\maketitle





\label{firstpage}


\begin{abstract}
Microarray studies, in order to identify genes associated with an outcome of interest, usually produce noisy measurements for a large number of gene expression features from a small number of subjects. One common approach to analyzing such high-dimensional data is to use linear errors-in-variables models; however, current methods for fitting such models are computationally expensive. In this paper, we present two efficient screening procedures, namely corrected penalized marginal screening and corrected sure independence screening, to reduce the number of variables for final model building. Both screening procedures are based on fitting corrected marginal regression models relating the outcome to each contaminated covariate separately, which can be computed efficiently even with a large number of features. Under mild conditions, we show that these procedures achieve screening consistency and reduce the number of features considerably, even when the number of covariates grows exponentially with the sample size. Additionally, if the true covariates are weakly correlated, corrected penalized marginal screening can achieve full variable selection consistency. Through simulation studies and an analysis of gene expression data for bone mineral density of Norwegian women, we demonstrate that the two new screening procedures make estimation of linear errors-in-variables models computationally scalable in high dimensional settings, and improve finite sample estimation and selection performance compared with estimators that do not employ a screening stage. 
\end{abstract}

%



\maketitle


%

\section{Introduction}
\label{section:intro}
In microarray studies, to identify genes that are associated with an outcome of interest, a large number of gene expressions (potentially tens of thousands) are measured from typically a much smaller number of subjects (often in the tens to hundreds). The gene expression measurements tend to be noisy, where measurement errors come from many sources such as sample preparation, labeling, and hybridization; for example, see \citet{rocke2001model} and \cite{zakharkin2005sources}. The gene measurements are also often analyzed on the log scale, making the assumption of additive measurement errors more plausible \citep{nghiem2019simselex}. Furthermore, as in common genome wide association studies \citep[among others]{do2011web, zhou2018brain}, it is usually assumed that only a few genes are related to the outcome of interest, i.e a sparsity assumption on the statistical model. As a specific motivating example, after some preprocessing steps, our Bone Mineral Density data in Section \ref{section:application} contains noisy measurements of $p=993$ features (genes) from $n=84$ observations (Norwegian women), and we are interested in identifying genes that are associated with the total hip T-score.    


For analyzing such data, a commonly used approach which we focus on in this paper is the classical linear errors-in-variables (EIV) model 
\begin{equation}
	\mathbf{y}  = \bX\bm{\beta}_0 + \bm\varepsilon, \quad
	\bW  = \bX+{\bU},
\label{eq:linear-eiv}
\end{equation}
where $\mathbf{y} \in \mathbb{R}^n$ is a random vector of outcomes from $n$ independent and identically distributed (iid) observations, $\mathbf{X} \in \mathbb{R}^{n\times p}$ is the deterministic true covariate  matrix (typically the true gene expressions), $\bbeta_0 = (\beta_{01}, \ldots, \beta_{0p})^\top \in \mathbb{R}^{p}$ is the true coefficient vector, and $\bm\varepsilon \in \mathbb{R}^{n}$  is the model error vector term whose components are assumed to be iid with zero mean and variance $\sigma^2$. Due to the existence of measurement error, the true covariate matrix $\bX$ is not observed; instead, we observe the random matrix $\mathbf{W} \in \mathbb{R}^{n\times p}$, which is a noisy version of $\mathbf{X}$, contaminated by an additive random measurement error matrix $\mathbf{U} \in  \mathbb{R}^{n\times p}$ independent of $\bm\varepsilon$. The rows of $\mathbf{U}$ are assumed to be iid random vectors with zero mean and covariance matrix $\bm{\Sigma}_{u}$. We focus on the model \eqref{eq:linear-eiv} in high dimensional settings, where the number of covariates $p$ can be bigger than and grow with the sample size $n$, potentially at an exponential rate. We also assume the true coefficient vector $\bbeta_0$ to be sparse, meaning that only a few components of $\bbeta_0$ are non-zero. 

When the true covariate matrix $\bX$ is observed, penalized regression methods \citep[among others]{tibshirani1996regression,fan2004nonconcave, huang2008asymptotic, simon2013sparse, piironen2017sparsity, ida2019fast} are widely used to estimate and perform variable selection on $\bbeta_0$. However, when the true covariate matrix $\mathbf{X}$ is not observed, replacing $\mathbf{X}$ with $\mathbf{W}$ leads to a naive estimator that is inconsistent in both estimation and variable selection of $\bbeta_0$ \citep{sorensen2015measurement}. To address this challenge, several corrections for measurement error in high dimensional linear EIV models have been proposed. For example, \citet{rosenbaum2010sparse} and \citet{rosenbaum2013improved} proposed the matrix uncertainty (MU) selector and its improved version, respectively, while \cite{belloni2017linear} proved its near-optimal minimax properties and developed a conic programming estimator that can achieve the minimax bound. Both the MU selector and conic estimator require appropriate choice of multiple tuning parameters, which is typically very challenging in practice, especially in high dimensional settings. Another approach for handling measurement error is to modify the loss function or the conditional score functions commonly seen in the error-free penalized regressions; examples include the corrected lasso method of \citet{loh2012} and \citet{sorensen2015measurement}, and the convex conditioned lasso of \citet{datta2017cocolasso}. More recently, \citet{romeo2019model} presented a simulation study to compare the performance of the MU, corrected lasso, and convex conditioned lasso against the naive estimator in both estimation and variable selection; they concluded that the relative performance of those estimators depend on the structure of $\bm\Sigma_u$. \citet{brown2019meboost} introduced a boosting algorithm based on the estimating equation of the corrected lasso, but the theoretical properties of the final estimates were not examined. In another line of research, \citet{nghiem2019simselex} proposed a SIMSELEX estimator that first uses simulation to evaluate the effect of measurement error on estimated coefficients, then selects important covariates based on these simulated effects, and finally extrapolates the simulation to the scenario with no measurement error present. Furthermore, \citet{byrd2019simple} presented an EM-type correction method, where they iteratively sample the true covariate from the conditional distribution of $\bX$ given $\mathbf{y}$ and $\bW$, and fitted penalized regressions of $\mathbf{y}$ on these sampled covariates. A major disadvantage of  all these methods is that they are not computationally efficient when the number of covariates $p$ is very large. Specifically, while the corrected lasso is defined to be a global minimum of a non-convex optimization problem, the convex conditioned lasso requires computation of the nearest semi-positive definite matrix measured in element-wise max norm in high dimensions; we will elaborate these two estimators in Section \ref{section:sim_est_metrics}. In addition, the SIMSELEX procedure requires running the lasso on a large number of simulated datasets, while the method of \citet{byrd2019simple} requires sampling from a large $p$-dimensional multivariate distribution, which is slow in high dimensional settings. 

In this paper, motivated by the Bone Mineral Density data, we address the large $p$ problem for linear EIV models by proposing two efficient corrected marginal screening methods, namely \emph{corrected penalized marginal screening} and \emph{corrected sure independence screening}, respectively. These screening methods aim to quickly identify a screening index set $\hat{\mathcal{Q}} \subset \{1, \ldots, p \}$ from the observed data, such that $\hat{\mathcal{Q}}$ has much smaller cardinality than $p$ but still retains all the important covariates, a property known as screening consistency. When no measurement error is present, screening methods in high dimensions are usually carried out based on marginal regressions, with the first two proposed methods being the penalized marginal bridge regression of \citet{huang2008asymptotic} and sure independence screening of \citet{fan2008sure} for the linear model, followed by a vast literature that improved these two methods and applied them to more complex models, see \citet[among others]{fan2008sure,fan2010sure, li2012robust, barut2016conditional, wen2018sure}. Compared to penalized methods when no measurement error is present, screening methods have received far less attention for EIV models. A screening method for linear EIV models was briefly mentioned in \citet{kaul2016two}. As detailed in Section \ref{section: methodology}, this method is a special case of our proposed corrected sure independence screening method when all the covariates have the same measurement error variance. In our new methods, after screening, we only compute penalized estimators using the variables indexed by $\hat{\mathcal{Q}}$; since the cardinality of the set $\hat{\mathcal{Q}}$ is much smaller than $p$, the total estimation times of two-stage estimators are much reduced compared to those of one-stage estimators that do not employ a screening. We demonstrate that the benefit of the proposed screening procedures is so substantial that it can make many estimators (such as the convex conditioned lasso) computationally feasible in high dimensional settings. Theoretically, we show that under mild conditions, even when the number of covariates $p$ grows at an exponential rate with the sample size $n$, our proposed screening methods achieve screening consistency while reducing the number of variables to below the sample size. Moreover, under a partial orthogonality condition for the true covariates, we demonstrate that the corrected penalized marginal screening approach can achieve full variable selection consistency. Our simulation studies and an analysis of the motivating Bone Mineral Density data both verify our theoretical results, and demonstrate that our proposed screening procedures lead to remarkable gains in both computational cost and finite sample performance. 

The remainder of this paper is organized as follows. Section \ref{section: methodology} introduces the proposed screening procedures and establishes their theoretical properties. In Section \ref{section:simulation}, we present simulation studies to demonstrate the strong empirical performance of screening procedures and several two-stage estimators. Section \ref{section:application} applies the methodologies to analyze the motivating Bone Mineral Density data, and Section \ref{section:conclusion} offers some concluding remarks. 

The following notation is used throughout the paper. For a generic matrix $\mathbf{A}$, let $A_{ij}$ denote the $(i,j)$ element of $\mathbf{A}$ and let $\lVert\mathbf{A}\rVert_2$, $\lVert\mathbf{A}\rVert_F$, and $\lVert\mathbf{A}\rVert_\text{max}$  denote the $\ell_2$ norm, Frobenius norm, and element-wise max norm, respectively. For a square matrix $\mathbf{A}$, let $\lambda_\text{max}(\mathbf{A})$ denote its maximum eigenvalue.  For any vector $\mathbf{v}$, let $v_j$ denote its $j$th component, $\lVert\mathbf{v}\rVert_2$ and $\lVert\mathbf{v}\rVert_1$ denote its $\ell_2$ norm and $\ell_1$ norm, respectively. For any set $\mathcal{S}$, let $\mathcal{S}^c$ denote its complement, and $\vert \mathcal{S} \vert$ denote its cardinality.  Finally, for any sequence $a_n$ and $b_n$, we write $a_n \sim b_n$ if there exist positive constants $c_1$ and $c_2$ such that $c_1 a_n \leq b_n \leq c_2 a_n$.
   
\section{Corrected marginal screening procedures}
\label{section: methodology}
Consider the linear EIV model \eqref{eq:linear-eiv}, with the observed data  consisting of the outcome vector $\mathbf{y} \in \mathbb{R}^{n}$ and the surrogate matrix $\bW \in \mathbb{R}^{n \times p}$.  
Let $\sigma_1^2, \ldots, \sigma_p^2$ denote the diagonal elements of $\bm{\Sigma}_{u}$; if any true covariate $X_{ij}$ is measured without error, then the corresponding element $\sigma_j^2$ is set to 
zero.  We assume only $s$ components of $\bbeta_0$ are non-zero, where $s \ll \min(n, p)$, and without loss of generality, let $\mathcal{S}=\left\{1, \ldots, s\right\}$ and $\mathcal{S}^{c}=\left\{s+1, \ldots, p\right\}$ denote the set of indices corresponding to non-zero and zero components of $\bm\beta_0$, respectively. For the remainder of this paper, unless otherwise stated, we assume the covariance matrix $\bm{\Sigma}_{u}$ is known; in practice, $\bm{\Sigma}_{u}$ is usually estimated from replicate data \citep{carroll2006measurement}. For example, in our microarray data analysis in Section \ref{section:application}, replications are available in the form of multiple probes for each gene expression, and we follow a common procedure to estimate $\bm\Sigma_u$ from these data (further details are presented in Appendix B). 

We develop screening procedures that can reduce the number of covariates for model \eqref{eq:linear-eiv} but still maintain all the important variables. Moreover, these procedures are designed to be computationally scalable to high dimensional settings with $p\gg n$. Specifically, we propose to screen variables by  minimizing
\begin{equation}
L(\bm\beta) = \sum_{j=1}^{p} L_j(\beta_j) =   \frac{1}{n} \sum_{j=1}^{p} \left\{ \sum_{i=1}^{n} \left(y_i - W_{ij}\beta_j\right)^2 - {\sigma}_j^2\beta_j^2+ \lambda_n p_n(\vert \beta_j\vert) \right\},
\label{eq:objective}
\end{equation}
with $\lambda_n$ being a non-negative tuning parameter and $p_n(\vert \beta_j\vert)$ a penalty function on $\vert \beta_j\vert$. 
Let $\hat{\bm\beta} = \argmin_{\bm\beta} L(\bbeta)$. 
The function  $L(\bm\beta)$ to be minimized in \eqref{eq:objective} consists of two parts. The first part, $\sum_{j=1}^{p} n^{-1}  \sum_{i=1}^{n} \left(y_i - W_{ij}\beta_j\right)^2 - \sum_{j=1}^{p}{\sigma}_j^2\beta_j^2$, is the sum of all the $\ell_2$ losses for the regression of the outcome on each surrogate predictor (i.e column of $\bW$) separately, with measurement error accounted for via a (negative) $\ell_2$ penalty term, $-\sum_{j=1}^{p}\sigma_j^2 \beta_j^2$. When $p$ is fixed and $n \to \infty$, this loss part converges to $n^{-1} \sum_{j=1}^{p}  \sum_{i=1}^{n} \left(y_i - X_{ij}\beta_j\right)^2$, which is the loss that was used by both \citet{huang2008asymptotic} and \citet{fan2010sure} for their corresponding marginal screening procedures when no measurement error is present. In the theoretical analysis below, we allow $p$ to diverge to $\infty$; in this case, this loss part still enables us to achieve screening consistency. For the second part of \eqref{eq:objective}, the penalty function $\lambda_n \sum_{j=1}^{p}p_n(\vert \beta_j\vert)$ regularizes the estimates, i.e with appropriate choices for $\lambda_n$, many components of the estimated $\hat{\bm{\beta}}$ are set to zero. Common choices of the penalty function $p_n(\vert\beta_j\vert)$ include the lasso $\vert \beta_j \vert$, the bridge penalty $\vert \beta_j \vert^\alpha$ with $0 < \alpha < 1$ \citep{frank1993statistical}, and the SCAD penalty  \citep{fan2004nonconcave}, among others. In this paper, we use the bridge penalty, which makes the solution of \eqref{eq:objective} relatively fast to compute and easy to analyze theoretically. However, we note that other penalties can be used in practice, although pursuing this is beyond the scope of this paper. Computationally, for any value of $\lambda_n$, each component $\hat{\beta}_j$ can be obtained by minimizing $L_j(\beta_j)$ separately. When $n$ is large enough, the coefficient $n^{-1}\sum_{i=1}^{n} W_{ij}^2 - \sigma_j^2$ associated with $\beta_j^2$ is positive, so $L_j(\beta_j)$ is a convex function of $\beta_j$ and can be minimized quickly using univariate convex optimization routines. Therefore, the minimizer of \eqref{eq:objective} can be computed efficiently even with very large $p$, making it and subsequent screening methods scalable in high dimensions.     

We consider two screening approaches arising from \eqref{eq:objective}. For the first approach, which will be referred to as \emph{corrected penalized marginal screening} (PMSc), we consider the case when the tuning parameter $\lambda_n$ is strictly positive. In this case, although the minimizer of \eqref{eq:objective} does not generally have a closed form, some components of $\hat{\bm\beta}$ can be set exactly to zero, so the corresponding screening set is defined to be $\hat{\mathcal{Q}}_\text{PMSc} = \{j : \hat{\beta}_j \neq 0\}$. For the second approach, we consider the case when  $\lambda_n = 0$, so the minimizer of \eqref{eq:objective} has components 
\begin{equation}
\tilde\beta_j = \dfrac{\sum_{i=1}^{n}W_{ij}y_i}{\sum_{i=1}^{n} W_{ij}^2- n\sigma_j^2}, \quad j =1,\ldots, p,
\label{eq:SISccomponent}
\end{equation}
each of which is non-zero with probability one. Note that each component $\tilde{\beta}_j$ is a consistent estimator of the slope $\gamma_j$ in the marginal univariate model between the outcome and the true (unobserved) covariate  $y_i = \gamma_j X_{ij} + \varepsilon_i, ~i=1,\ldots, n$, which is the quantity that is used by \citet{fan2008sure} and \citet{fan2010sure} in sure independence screening (SIS) when no measurement error exists. In order to reduce the number of dimensions in this unpenalized approach, similar to SIS, we keep the $d$ components with largest magnitude $\vert \tilde{\beta}_j \vert, j =1,\ldots, p$, and refer to this approach as \emph{corrected sure independence screening} (SISc). The corresponding screening set is then defined to be
\begin{equation}
\hat{\mathcal{Q}}_\text{SISc} = \{1 \leq j \leq p: \vert \tilde{\beta}_j \vert \text{ is among the first}~ d~ \text{largest of all components} ~ \vert\tilde{\beta}_1 \vert, \ldots, \vert \tilde{\beta}_p \vert \}. 
\label{eq:setQ0}
\end{equation}
We remark that the SISc approach in our paper is more general than the corrected screening method introduced by \citet{kaul2016two} in the first step of their two-step estimation procedure; specifically, \citet{kaul2016two} screened the variables based on $\zeta_j =\sum_{i=1}^{n} W_{ij} y_i, ~ j =1,\ldots, p$, which is the numerator of \eqref{eq:SISccomponent}. The rank of $\vert\zeta_j\vert$ is asymptotically the same as the rank of $\vert\tilde\beta_j\vert$ if all the covariates have the same measurement error variance and are measured on the same scale. Therefore, the method of \citet{kaul2016two} may be considered as a special case of SISc when all the covariates have the same measurement error variances. 

In the next two subsections, we study the theoretical properties of the two screening procedures PMSc and SISc separately; all the proofs are provided in Appendix A. Throughout the development below, we note that the true covariate matrix $\bX$ is assumed to be deterministic; this assumption could be relaxed to allow $\bX$ to be random, although we do not explore this in the paper. 

\subsection{Corrected penalized marginal bridge screening}
First, we study the properties of the PMSc approach with the penalty function $p_n(\vert \beta_j \vert) = \vert\beta_j\vert^\alpha$ for $0 < \alpha < 1$. Since the screening step aims only to reduce the dimension of the feature space, we are interested only in variable selection properties, noting that $\hat\bbeta$ is generally not estimation consistent for the true vector $\bbeta_0$. For the theoretical development below, we define 
$$
\tilde{\xi}_{n j}= \frac{1}{n} \mathbb{E}\left(\sum_{i=1}^{n} y_{i} X_{ij}\right) = \frac{1}{n} \sum_{i=1}^{n}\left(\sum_{k=1}^{s} X_{ik}{\beta}_{0k}\right) X_{i j},
$$
$$
\xi_{n j}= \frac{1}{n} \mathbb{E}\left[\left(\sum_{i=1}^{n} y_{i} W_{ij}\right) \bigg\rvert ~U_{ij}\right]= \frac{1}{n} \sum_{i=1}^{n}\left(\sum_{k=1}^{s} X_{ik}{\beta}_{0k}\right) W_{i j} = \tilde{\xi}_{n j} + \frac{1}{n}\sum_{i=1}^{n}\left(\sum_{k=1}^{s} X_{ik}{\beta}_{0k}\right) U_{ij}.
$$
Here, $\tilde{\xi}_{nj}$ can be considered as the marginal covariance between the outcome and the $j$th true covariate. As in the no measurement error case, this quantity $\tilde\xi_{nj}$ with $j \in \mathcal{S}$ will play an essential role in determining whether the component $\hat{\beta}_j$ is asymptotically non-zero. However, because the $X_{ij}$'s are not observed, the quantities $\tilde{\xi}_{nj}$ are not directly computable, so the random variables $\xi_{nj}$ are used as surrogates for $\tilde{\xi}_{nj}$, $j=1,\ldots, p.$ Next, we assume the following conditions.

\begin{enumerate}[label=(C\arabic*),leftmargin =*, ref=C\theenumi]	   

\item  \label{cond:epsilonsubGaussian} The model error terms $\varepsilon_{1}, \varepsilon_{2}, \ldots \varepsilon_n$ are iid sub-Gaussian random variables with mean zero and finite variance $\sigma^{2}$, 
 \label{cond:B1}

\item \label{cond: MEsubGaussian} The measurement errors $U_{ij},~i=1,\ldots,n$ are independent sub-Gaussian random variables with finite variance $\sigma_j^2$, $j = 1,\ldots, p$. 
 \label{cond:A2} 

\item  \label{cond:B4} \label{cond:boundonbetak} There exist constants $b_0$ and $b_1$ such that $\displaystyle 0 < b_0 \leq \min _{k \in \mathcal{S}}\left|\beta_{0 k}\right| \leq \max _{k \in \mathcal{S}}\left|\beta_{0 k}\right| \leq b_{1} < \infty. $ 

\item \label{cond:B6} There exist constants $C_1$ and $C_2$ such that 
$\displaystyle 0 < C_1 \leq \min_{j} n^{-1} \sum_{i=1}^{n} X_{ij}^2 < \max_{j} n^{-1}\sum_{i=1}^{n} X_{ij}^2 \leq C_2 < \infty. \label{cond:boundonxsquare} 
$

\item \label{cond:new} There exists a constant $\xi_0$ such that $\min _{k \in \mathcal{S}}|\tilde{\xi}_{n j}|\geq \xi_{0}>0 $ for all $n$. \label{cond:boundonxandy}
\end{enumerate}

Conditions \eqref{cond:epsilonsubGaussian}-\eqref{cond:A2} are standard in the high dimensional measurement error literature; e.g  these conditions were used in \citet{rosenbaum2013improved} and \citet{belloni2017linear}. Condition \eqref{cond:B4} requires that all the true non-zero components of $\bbeta_0$ are bounded away from zero and infinity. Condition \eqref{cond:B6} assumes that each true covariate is well-controlled; in particular, the lower bound implies that the ``signal'' part on each covariate is not too small (for example, the true covariate matrix $\bX$ should not be too sparse). With the above conditions, the PMSc procedure can achieve screening consistency, as made precise in Theorem \ref{theorem:screening_consistency}.
\begin{theorem}
	Consider the linear EIV model \eqref{eq:linear-eiv} under conditions \eqref{cond:epsilonsubGaussian}-\eqref{cond:new}, and the screening set $\hat{\mathcal{Q}}_{\text{PMSc}} = \{j : \hat\beta_j \neq 0\}$, where $\hat{\bm\beta}$ minimizes \eqref{eq:objective} with the tuning parameter $\lambda_n \rightarrow 0$ and the number of important variables $s = o(n)$. Then, we have
	$
	\mathbb{P}(\hat{\mathcal{Q}}_{\text{PMSc}} \supseteq\mathcal{S} ) \rightarrow 1,
	$ meaning all the important variables are included in the screening set with probability tending to one as $ n \to \infty$.
	\label{theorem:screening_consistency}
\end{theorem}
The conditions for screening consistency as stated in Theorem \ref{theorem:screening_consistency} are relatively weak. Indeed, it is easy to see that setting the tuning parameter $\lambda_n = 0$, such as in the case of SISc, always guarantees screening consistency, although no dimension reduction is achieved in this case. If the true important and non-important covariates are weakly correlated, as formalized in the condition \eqref{cond:B2} below, then the PMSc procedure can achieve full selection consistency, meaning that the screening set contains important variables only. Specifically, we assume the following additional conditions. 
\begin{enumerate}[label=(C\arabic*), leftmargin = *, ref = C\theenumi]
	\setcounter{enumi}{5}	
	\item \label{cond:B2} There exists constants $d_{0}>0 $ and $0 \leq \theta\leq 1/2$ such that
	$\displaystyle
	\left|n^{-1} \sum_{i=1}^{n} X_{i j} X_{i k}\right| \leq d_{0} n^{-\theta}, ~~j \in \mathcal{S}^c,~ k \in \mathcal{S}.
	$
	\item \label{cond:lambda} Assume the tuning parameter $\lambda_n$ satisfies:
	\begin{enumerate}[label = (\alph*), ref=\theenumi\alph*]
		\item $\lambda_n \to 0$ and  $\lambda_{n} n^{\theta(2-\alpha) } s^{\alpha-2} \rightarrow \infty$, \label{cond:C9a}
		\item  $\log \left(2m\right)= o(1) \times \left\{\lambda_{n} n^{(2-\alpha) / 2}\right\}^{1 /(2-\alpha)},$ where $ m = p-s$. \label{cond:C9b}
	\end{enumerate}
\end{enumerate}
Condition \eqref{cond:B2} is similar to the partial orthogonality condition of \citet{huang2008asymptotic}, which is necessary to establish the full variable selection consistency of the penalized marginal bridge estimator when no measurement error is present. Such a condition is also necessary when measurement errors are present. Condition \eqref{cond:C9a} implies that the number of important variables $s$ is of order $o(n^{\theta})$ (hence is of order $o(n^{1/2})$ at most), while condition \eqref{cond:C9b} implies the number of zero coefficients $m = p-s$ can be of order $o(\exp(n^{1/2}))$. As a result, the number of variables $p$ is allowed to grow at an exponential rate with $n^{1/2}$, while the number of important variables $s$ grows at a comparably slower rate. These two conditions also control the behavior of the tuning parameter $\lambda_n$ that should be used for PMSc to achieve full variable selection consistency.
\begin{theorem}
\label{theorem:full_consistency}
    Consider the linear EIV model \eqref{eq:linear-eiv} under conditions \eqref{cond:epsilonsubGaussian}-\eqref{cond:lambda}, and the screening set $\hat{\mathcal{Q}}_{\text{PMSc}} = \{j : \hat\beta_j \neq 0\}$. Then we have
$
\mathbb{P}(\hat{\mathcal{Q}}_\text{PMSc} = \mathcal{S}) \rightarrow 1,
$ 
meaning the screening set $\hat{\mathcal{Q}}_\text{PMSc}$ contains all and only the important variables as $n \to \infty$.
\end{theorem}
Note that in both Theorem \ref{theorem:screening_consistency} and Theorem \ref{theorem:full_consistency}, we do not make any assumption on the structure of the covariance matrix $\bm\Sigma_u$, except that the measurement error variance on each covariate is bounded for all $j=1,\ldots, p$, as formalized in condition \eqref{cond: MEsubGaussian}. In Appendix C, we demonstrate empirically that the conclusion from Theorem \ref{theorem:full_consistency} holds even when the measurement error among the covariates is highly correlated. 

In practice, since the true covariate matrix $\bX$ is not observed, condition \eqref{cond:B2} is hard to verify, and even when it is assumed to hold, we stress that the theoretical choice of $\lambda_n$ in condition \eqref{cond:lambda} depends on unknown quantities. Therefore, in practice, for the proposed PMSc procedure, we use cross-validation to select the tuning parameter $\lambda_n$, where the loss on test data is computed based on the loss part of \eqref{eq:objective}. This procedure will be referred to as PMSc cross-validation  (PMSc$_\text{CV}$) and is further elaborated upon in Subsection \ref{section:sim_est_metrics}.  Alternatively, we can start with a sufficiently large value for $\lambda_n$ such that all the coefficients are set to zero, then incrementally decrease $\lambda_n$ until a certain number of covariates, $M$, is included. Note that with this version of PMSc, rather than having to choose the tuning parameter $\lambda_n$, we instead choose $M$, the desired number of covariates to keep from the screening procedure. This strategy will be referred to as PMSc forward stepwise (PMSc$_\text{FS}$). As demonstrated in Appendix C, when the true model is sparse as is assumed in our high dimensional settings, in order to retain all the important variables, $M$ can be substantially smaller than the number of covariates $d$ to keep in the SISc approach, and hence also be substantially smaller than the sample size $n$. 

\subsection{Corrected sure independence screening}
Because SISc corresponds to the minimizer of \eqref{eq:objective} with $\lambda_n = 0$, it screens out important variables by keeping only the $d$ components with largest magnitude among $\{\vert\tilde{\beta}_1\vert, \ldots, \vert\tilde{\beta}_p\vert\}$ as formalized by $\hat{\mathcal{Q}}_\text{SISc}$ in \eqref{eq:setQ0}. 
In this section, we prove that under certain conditions, the scalar $d$ can be chosen such that we can reduce the number of dimensions to below the sample size $n$ and still retain all the important variables. To achieve this, we follow the theoretical development of SIS in the linear model with no measurement error by \citet{fan2008sure}, and aim to theoretically set $d = \floor{\gamma n}$, where $\gamma$ can be of order $n^{-\tilde{\theta}}$ for some $0 < \tilde\theta < 1$. In addition to the conditions \eqref{cond:epsilonsubGaussian}-\eqref{cond:boundonxandy} for screening consistency as in Theorem \ref{theorem:screening_consistency}, we impose the following additional assumptions.
\begin{enumerate}[label= (C\arabic*), leftmargin =*, ref=C\theenumi]	 
	\setcounter{enumi}{7}	
	\item $p >n$, $\log(p)/n \to C_1$ with $C_1$ being a constant as both $n$ and $p$ diverge to $\infty$.	\label{cond:np}
	\item  Let $\mathbf{Z} = \bU \bm\Sigma_u^{-1/2}$. Then, 
	\begin{enumerate}[label = (\alph*), ref=\theenumi\alph*]
		\item \label{cond:MEconcentration} For some  constants $c_1, c_2 >1$ and $D> 0$, the matrix  $\mathbf{Z}$ follows a spherical distribution satisfying
		$
			\mathbb{P}\left\{\lambda_{\max }\left(\tilde{p}^{-1} \tilde{\mathbf{Z}} \widetilde{\mathbf{Z}}^{\top}\right)>c_{2} \text { and } \lambda_{\min }\left(\widetilde{p}^{-1} \tilde{\mathbf{Z}} \tilde{\mathbf{Z}}^{\top}\right)<1 / c_{2}\right\} \leq e^{-Dn}, ~
		$
		for any $ n\times \tilde{p}$ submatrix $\tilde{\mathbf{Z}}$ of $\mathbf{Z}$ with $c_1n \leq \tilde{p} \leq p$.
		\item \label{cond:MEeigenvalue} There exists positive constants $\tau_1>0$ and $c_3>0$, such that 
		$
		\lambda_{\max }(\bm{\Sigma}_u) \leq c_3n^{\tau_1}.
		$
	\end{enumerate}

	\item \label{cond:boundoneigX} There exists positive constants $c_4>0$ and  $\tau_2 > 0$, such that
		$
		\lambda_\text{max}\left(n^{-1}\bX^\top \bX\right) \leq c_4n^{\tau_2}.
		$
	\item \label{cond:tau} $\tau_1 + \tau_2 + \log_n(s)  <1$, where  $\log_n(s)$ denotes the logarithm base $n$ of $s$. 
\end{enumerate}
Condition \eqref{cond:np} implies that we can allow the number of covariates $p$ to grow exponentially with the sample size $n$. Condition \eqref{cond:MEconcentration} is referred to as the Concentration Property by \citet{fan2008sure}, meaning that with large probability, the $n$ nonzero singular values of the $n \times \tilde{p}$ submatrix $\tilde{\mathbf{Z}}$ of $\mathbf{Z}$ are of the same order; \citet{fan2008sure} suggested that a sufficient condition for \eqref{cond:MEconcentration} is that each row $\mathbf{U}_i$ follows a $p$-variate Gaussian distribution. Conditions \eqref{cond:MEeigenvalue} and \eqref{cond:boundoneigX} indicate that the maximum eigenvalues of the covariance measurement error matrix $\bm{\Sigma}_u$ and of the scaled Gram matrix $n^{-1} \bX^\top \bX$ can only grow polynominally with the sample size $n$. Furthermore, condition \eqref{cond:tau} restricts the degree of that polynomial growth to be less than one, which rules out the case when either true covariates or measurement errors are highly correlated.  Condition \eqref{cond:tau} also implies that the number of important covariates $s$ is smaller than the sample size $n$, so that reducing the dimension to below $n$ is reasonable. Compared with the theoretical conditions used to establish screening and full variable selection consistency of PMSc (Theorems \ref{theorem:screening_consistency} and \ref{theorem:full_consistency}), conditions \eqref{cond:np}-\eqref{cond:tau} impose stricter conditions on the distribution and covariance matrix of measurement error, but allow the number of covariates $p$ to grow at a faster rate. 

With the conditions above, we establish the following theorem regarding screening consistency of the proposed SISc procedure. 

\begin{theorem}
\label{theorem:SISc}
Consider the linear EIV model \eqref{eq:linear-eiv} under conditions \eqref{cond:epsilonsubGaussian}-\eqref{cond:boundonxandy} and \eqref{cond:np}-\eqref{cond:tau}, and the screening set $\hat{\mathcal{Q}}_{\text{SISc}}$ defined in \eqref{eq:setQ0} with $d = \floor{\gamma n}$ and $\tilde\beta_j$ given by \eqref{eq:SISccomponent}. Then there exists some $\tilde\theta < 1-\tau_1-\tau_2-\log_n(s)$ such that when $\gamma \sim cn^{-\tilde\theta}$ with $c>0$, we have 
$
\mathbb{P}\left(\hat{\mathcal{Q}}_{\text{SISc}} \supseteq \mathcal{S} \right)=1-O\left(p\exp \left(-C n \right)\right)
$ for some positive constant $C$.
\end{theorem}

Theorem \ref{theorem:SISc} implies that the SISc procedure can reduce exponentially high dimension to a relatively small dimension $d = O(n^{1-\tilde\theta}) < n$ while retaining all the important covariates. In practice, as suggested by \citet{fan2008sure}, common choices for $d$ include $d = \floor{n/\log(n)}$ or $d = n - 1$. Finally, unlike PMSc, a screening procedure based on ranking components such as SISc can only achieve full selection consistency if we know the true number of important covariates $s$ in advance. Since that is rarely the case in practice, then we do not study the theoretical conditions under which we can choose $d=s$. 

\section{Simulation studies}
\label{section:simulation}

\subsection{Simulation setup}
\label{sim1}

We conducted simulation studies to demonstrate the benefit of the proposed screening procedures by comparing the performance of two-stage estimators, which screen variables in the first stage and compute a corrected penalized estimator on the retained variables in the second stage, against one-stage estimators, which do not employ any screening method. We simulated data from model \eqref{eq:linear-eiv}, where each row of the true covariate matrix $\bX_i$ was generated from a $p$-variate Gaussian distribution with a zero mean vector and two choices for the covariance matrix $\bm{\Sigma}_x$. In the first scenario, $\bm{\Sigma}_x$ had an autoregressive AR(1) structure with elements $\sigma_{ij} = \rho_x^{\vert i -j \vert}$. In the second scenario, $\bm{\Sigma}_x$ had an homogeneous (exchangeable) structure where all diagonal and off-diagonal elements were set to 1 and $\rho_x$ correspondingly. For both scenarios, we varied $\rho_x \in \{0.3, 0.5\}$. The true $p$-dimensional vector $\bm\beta_0$ was constructed such that the first $s=5$ non-zero components were generated from the uniform distribution $U(1,1.5)$. Turning to the measurement errors, we generated $\bU_i$ independently of $\bX_i$ from another $p$-variate Gaussian distribution with a zero mean vector and two choices of the covariance matrix $\bm{\Sigma}_u$. For the first choice, $\bm{\Sigma}_u$ was a diagonal matrix with elements randomly generated from the uniform $U(0.1, 0.5)$ distribution; hence, the measurement errors on each covariate were independent of those of the other covariates with the noise-to-signal ratio ranging from 10\% to 50\%. For the second choice, $\bm{\Sigma}_u$ had a block diagonal structure, where the $p$ covariates were divided into $p/4$ non-overlapping groups of size $4$ with the correlation between any pair in the same group equal to $0.2$, and the diagonal elements of $\bm{\Sigma}_u$ equal to $0.4$; hence, the measurement errors on one covariate were positively correlated with those on three other covariates with the noise-to-signal ratio being $40\%$. We set the sample size to $n=500$, and varied the number of covariates $p \in \{1000, 2000\}$. Finally, the elements of the model error term $\bm{\varepsilon}$ were independently generated from a Gaussian distribution with mean zero and variance $\sigma^2 = 0.25.$

\subsection{Estimators and performance metrics}
\label{section:sim_est_metrics}
We computed several one-stage and two-stage estimators on each simulated dataset. For one-stage estimators, we implemented the corrected lasso estimator of \citet{loh2012} and \citet{sorensen2015measurement}, and the convex conditioned (CoCo) lasso of \citet{datta2017cocolasso}. Without using any screening procedure, the one-stage corrected lasso is given by
\begin{equation}
\hat{\bm\beta}^\text{CL} = \argmin_{\bbeta} \left\{ \dfrac{1}{n} \sum_{i=1}^{n}(y_i - \bW_i^\top \bm\beta)^2 - \bm\beta^\top \bm\Sigma_u \bm\beta + \mu \lVert\bm\beta\rVert_1 \right\}, ~\text{subject to ~ } \lVert\bm\beta\rVert_1 \leq R
\label{eq:correctlasso},
\end{equation}
with $\mu$ and $R$ being two positive tuning parameters. As noted in \citet{loh2012}, the problem  \eqref{eq:correctlasso} is generally non-convex when $p > n$, because the matrix $n^{-1} \sum_{i=1}^{n}\bW_i^\top \bW_i -\bm{\Sigma}_u$ has a large number of negative eigenvalues and is not positive definite. Therefore, classic gradient descent algorithms, as outlined in \citet{loh2012} and \cite{sorensen2015measurement}, are only guaranteed to converge to a local minimum under a careful choice of the tuning parameters $\mu$ and $R$. \cite{sorensen2015measurement} suggested that the tuning parameter $\mu$ be chosen as the minimum of the ten-fold cross-validation curve of the naive lasso
$$
\hat{\bm\beta}^\text{naive} = \argmin_{\bm\beta} \left\{\dfrac{1}{n} \sum_{i=1}^{n}(y_i - \bW_i^\top \bm\beta)^2  + \mu \lVert\bm\beta \rVert_1 \right\}, 
$$
and $R$ be chosen by another ten-fold cross-validation from a grid of equally spaced values between $10^{-3}\kappa$ and $\kappa$, with $\kappa = 2\lVert\hat{\bm{\beta}}^{\text{naive}}\rVert_1$.  The one-stage CoCo lasso estimator is given by  
$$
\hat{\bm\beta}^\text{CoCo}=\underset{\bbeta}{\argmin }\left\{\dfrac{1}{2} \bbeta^{\prime} \widetilde{\bm\Sigma} \bbeta-\tilde{\rho}^{\top} \bbeta+\tilde{\mu}\|\bbeta\|_{1} \right\},
$$
where $\tilde{\rho} =  \bW^\top \bm{y}$ and 
$
\widetilde{\bm\Sigma} = \underset{\bm{\Sigma} \in \mathbb{R}^{p\times p}_+}{\argmin} \lVert \bm{\Sigma} - \widehat{\bm{\Sigma}}  \rVert_\text{max}, 
$
with $\widehat{\bm{\Sigma}} = n^{-1}\mathbf{W}^\top \mathbf{W} - \bm{\Sigma}_u$ and $\mathbb{R}^{p\times p}_+$ the set of $p\times p$ positive semi-definite matrices. In other words, $\widetilde{\bm{\Sigma}}$ is the nearest positive semi-definite matrix to $\hat{\bm{\Sigma}}$ measured by the element-wise max norm, which was computed by an alternating direction method of multipliers (ADMM) algorithm. We point out that this element-wise max norm makes computation of $\widetilde{\bm\Sigma}$ so expensive when $p$ is large that in our simulation, we only computed the CoCo lasso estimator if the number of variables was smaller than $1000$. The non-negative tuning parameter $\tilde\mu$ for the CoCo lasso estimator was selected via a corrected cross-validation procedure; see \citet{datta2017cocolasso} and \citet{datta2020note} for more details. 

For two-stage estimators, we first implemented either PMSc$_{\text{CV}}$, PMSc$_{\text{FS}}$, or SISc, and then applied either the corrected lasso or CoCo lasso to the covariates selected from the first stage. For the two versions of PMSc, we set $\alpha = 0.5$ as is commonly done when the bridge penalty is used in practice \citep{huang2008asymptotic, huang2009group, polson2014bayesian}. For PMSc$_{\text{CV}}$, we selected the tuning parameter $\lambda_n$ via the following five-fold cross-validation procedure: on each simulated dataset, the data were randomly split into 5 folds $\mathcal{F}_1, \ldots, \mathcal{F}_5$ of equal sizes; for the $k$th iteration, the fold $\mathcal{F}_k$ was left out to be the test set and PMSc was applied on the remaining 4 folds on a grid consisting of $40$ equally spaced values of the tuning parameter $\lambda_n$. Let $\hat{\bbeta}_\lambda^{(-k)} = (\hat{\beta}_{1,\lambda}^{(-k)}, \ldots, \hat{\beta}_{p,\lambda}^{(-k)})^\top$ denote the solution of \eqref{eq:objective} when $\mathcal{F}_k$ was left out. Then the final $\lambda_n$ was selected to minimize 
$$
 \sum_{k=1}^{5}  \left\{ \sum_{i\in \mathcal{F}_k} \sum_{j=1}^{p} \left(y_i - \hat{\beta}_{j,\lambda}^{(-k)}\tilde{W}_{ij}\right)^2 - \vert\mathcal{F}_k\vert \sum_{j=1}^{p}\left(\hat\beta_{j, \lambda}^{(-k)}\right)^2\sigma_j^2 \right\},
$$ where $\vert\mathcal{F}_k\vert$ denotes the cardinality of the fold $\mathcal{F}_k$. For the  PMSc$_\text{FS}$ and SISc, the number of covariates to be included and kept was chosen to be $M = d = \floor*{n/\log(n)}$, respectively. This choice of $d$ and $M$ follows a common practice for screening procedures in high dimensional settings \citep{fan2008sure,zhu2011model,cui2015model}. The two-stage estimator when PMSc$_\text{CV}$ and corrected lasso are used in the first and the second stage respectively is referred to as the PMSc$_\text{CV}$-Corrected estimator. Similar definitions hold for other estimators resulting from other combinations of the methods used in the first and second stage. Also, we refer to all the estimators that used SISc in the first stage as SISc-based estimators; similar definitions are applied to the estimators that used another screening procedure in the first stage. 

We report false positive rate (FPR) and false negative rate (FNR) of the screening procedures, and compare all the one-stage and two-stage estimators based on FPR, FNR, and $\ell_2$ estimation error (i.e the $\ell_2$ norm of the difference between an estimate and the true coefficient vector). All metrics were averaged across 500 simulations. We also report the mean total computation time for each estimator. All the reported times are based on implementation on Artemis, a high performance computing cluster at the University of Sydney where every sample was computed on a single core (Intel(R) Xeon(R) CPU E5-2697A v4 2.60GHz and $4$GB RAM). 

\subsection{Simulation results}

Tables \ref{tab:AR1Sigmax_0.5_short} and \ref{tab:HomogenSigmax_0.5_short} present the summary results for the simulations with $\rho_x = 0.5$. The results for other settings show similar trend; complete results, including the standard errors and separation of total computation time into the first and second stages for all the settings, can be found in Appendix D.

When only a few true covariates are correlated with one another (i.e $\bm{\Sigma}_x$ had an $\text{AR}(1)$ structure), Table \ref{tab:AR1Sigmax_0.5_short} shows that all the screening procedures (first stage) were able to keep all the important variables and reduce the number of dimensions effectively. The PMSc$_{\text{CV}}$ method had FPR and FNR closest to zero; as a result, after the second stage, the PMSc$_{\text{CV}}$-based estimators had the lowest estimation error. The PMSc$_{\text{FS}}$ and SISc screening procedures had zero FNR but some false positives, which were expected from further investigation of Figure 1 in Appendix C, where we demonstrate that the choice of $d$ and $M$ in this simulation was greater than the minimum number of variables that PMSc$_{\text{FS}}$ and SISc needed to include in order to retain all the important variables. However, after the second stage, the FPRs of the two-stage PMSc$_{\text{FS}}$ and SISc-based estimators were reduced to close to zero, and notably lower than the FPRs of the corresponding one-stage estimators; the gain was more pronounced when the CoCo lasso was used in the second stage. In turn, the estimation errors of the two-stage estimators were also lower than those of the one-stage estimators, with the improvement increasing when $p$ increased from $1000$ to $2000$. When $\bm\Sigma_u$ was diagonal, there was little difference in the $\ell_2$ error of the two-stage estimators; however, when $\bm{\Sigma}_u$ was block-diagonal, the SISc-based estimators tended to perform worse than the PMSc$_{\text{FS}}$ and PMSc$_{\text{CV}}$-based estimators. Regarding computation time, among the three screening methods, SISc was the fastest to compute, followed by PMSc$_{\text{FS}}$ and PMSc$_{\text{CV}}$. As a whole, the total computation times of the two-stage estimators were considerably lower than those of the one-stage estimators; further investigation of Table C.1 in Appendix C shows that the screening stage substantially decreased the time needed for the second stage estimation. Finally, when $p$ increased from $p=1000$ to $p=2000$, there was a remarkable increase in the computation time of the one-stage estimators; however, the corresponding increases for the two-stage estimators were small, reflecting their computational scalability.  
\begin{table}[ht]
	\centering
	\caption{Performance of the one-stage and two-stage estimators in the simulation study based on mean false positive rate (FPR, in percentage), false negative rate (FNR, in percentage), $\ell_2$ error and computation time (in seconds) when $\bm{\Sigma}_x$ has an AR(1) structure with autocorrelation $\rho_x = 0.5$. The CoCo estimator was only computed when the number of variables was no more than $1000$ (either in 1st or 2nd step). }
	\medskip
	\resizebox{0.85\textwidth}{!}{\begin{tabular}{lll rr rrr r}
			\toprule[1.5pt]
			$\bm\Sigma_u$  & $p$ & Estimator &  \multicolumn{2}{c}{1st step} &
			\multicolumn{3}{c}{2nd step} & Time  \\
			& & & FPR & FNR & FPR & FNR & $\ell_2$ & \\
			\cmidrule(lr){4-5} \cmidrule(lr){6-8} 
			
		Diagonal  & 1000 & One-stage Corrected &  - &  - &  2.2 &  0.0 & 0.35 &  49.7 \\ 
		&  & PMSc$_{\text{CV}}$-Corrected &  0.1 &  0.0 &  0.1 &  0.0 & 0.26 &  23.8 \\ 
		&  & PMSc$_{\text{FS}}$-Corrected &  7.5 &  0.0 &  0.0 &  0.0 & 0.28 &   7.0 \\ 
		&  & SISc-Corrected &  7.5 &  0.0 &  1.2 &  0.0 & 0.33 &   3.5 \\ 
		&  & One-stage CoCo &  - &  - &  7.6 &  0.0 & 0.47 & 516.8 \\ 
		&  & PMSc$_{\text{CV}}$-CoCo &  0.1 &  0.0 &  0.1 &  0.0 & 0.26 &  23.4 \\ 
		&  & PMSc$_{\text{FS}}$-CoCo &  7.5 &  0.0 &  0.0 &  0.0 & 0.27 &   6.1 \\ 
		&  & SISc-CoCo &  7.5 &  0.0 &  0.1 &  0.0 & 0.31 &   2.2 \\ 
		& 2000 & One-stage Corrected &  - &  - &  1.3 &  0.0 & 0.36 & 136.5 \\ 
		&  & PMSc$_{\text{CV}}$-Corrected &  0.0 &  0.0 &  0.0 &  0.0 & 0.26 &  38.2 \\ 
		&  & PMSc$_{\text{FS}}$-Corrected &  3.8 &  0.0 &  0.6 &  0.0 & 0.29 &  20.0 \\ 
		&  & SISc-Corrected &  3.8 &  0.0 &  0.6 &  0.0 & 0.33 &   4.5 \\ 
		&  & PMSc$_{\text{CV}}$-CoCo &  0.0 &  0.0 &  0.0 &  0.0 & 0.26 &  38.0 \\ 
		&  & PMSc$_{\text{FS}}$-CoCo &  3.8 &  0.0 &  0.1 &  0.0 & 0.27 &  18.0 \\ 
		&  & SISc-CoCo &  3.8 &  0.0 &  0.1 &  0.0 & 0.31 &   3.5 \\ \\
		
		\multirow{2}{1.2cm}{Block diagonal} & 1000 & One-stage Corrected &  - &  - &  1.8 &  0.0 & 0.62 &  27.8 \\ 
		&  & PMSc$_{\text{CV}}$-Corrected &  0.0 &  0.0 &  0.0 &  0.0 & 0.43 &  23.8 \\ 
		&  & PMSc$_{\text{FS}}$-Corrected &  7.5 &  0.0 &  0.0 &  0.0 & 0.48 &   7.2 \\ 
		&  & SISc-Corrected &  7.5 &  0.0 &  1.0 &  0.0 & 0.57 &   4.1 \\ 
		&  & One-stage CoCo &  - &  - &  7.4 &  0.0 & 0.74 & 417.4 \\ 
		&  & PMSc$_{\text{CV}}$-CoCo &  0.0 &  0.0 &  0.0 &  0.0 & 0.43 &  23.4 \\ 
		&  & PMSc$_{\text{FS}}$-CoCo &  7.5 &  0.0 &  0.0 &  0.0 & 0.46 &   6.2 \\ 
		&  & SISc-CoCo &  7.5 &  0.0 &  0.1 &  0.0 & 0.54 &   3.2 \\ 
		& 2000 & One-stage Corrected &  - &  - &  1.0 &  0.0 & 0.64 & 68.0 \\ 
		&  & PMSc$_{\text{CV}}$-Corrected &  0.0 &  0.0 &  0.0 &  0.0 & 0.44 & 38.2 \\ 
		&  & PMSc$_{\text{FS}}$-Corrected &  3.8 &  0.0 &  0.5 &  0.1 & 0.46 & 19.5 \\ 
		&  & SISc-Corrected &  3.8 &  0.0 &  0.5 &  0.0 & 0.57 &  5.1 \\ 
		&  & PMSc$_{\text{CV}}$-CoCo &  0.0 &  0.0 &  0.0 &  0.0 & 0.45 & 37.9 \\ 
		&  & PMSc$_{\text{FS}}$-CoCo &  3.8 &  0.0 &  0.0 &  0.0 & 0.46 & 18.4 \\ 
		&  & SISc-CoCo &  3.8 &  0.0 &  0.1 &  0.0 & 0.54 &  4.2 \\ 
		\bottomrule[1.5pt]
	\end{tabular}}
\label{tab:AR1Sigmax_0.5_short}
\end{table}

When all the true covariates were highly correlated with each other (i.e $\bm{\Sigma}_x$ had a homogeneous structure), Table \ref{tab:HomogenSigmax_0.5_short} shows that PMSc$_{\text{CV}}$ was not helpful in either improving performance or reducing computation time. For the first step, PMSc$_{\text{CV}}$ kept all the variables, meaning no dimension reduction was achieved. As a result, the two-stage PMSc$_{\text{CV}}$-based estimators had essentially the same performance as the corresponding one-stage estimators. Additionally, PMSc$_{\text{FS}}$ and SISc had similar and much lower FPRs than PMSc$_{\text{CV}}$, with the former tending to have lower FNR than the latter. As further demonstrated in Web Figure 1 of Appendix C, these FNRs were expected because the minimum number of variables SISc had to keep in order to retain all the important variables was much higher than that of PMSc$_{\text{FS}}$. After the second stage, the FPRs of all the PMSc-based (both versions) and SISc-based estimators were much reduced, while the corresponding FNRs were unchanged. Regarding estimation error, the two-stage PMSc$_{\text{FS}}$ and SISc-based estimators had noticeably lower estimation errors than the corresponding one-stage estimators, with the improvement being greater when $p$ increased from $1000$ to $2000$. When $\bm{\Sigma}_u$ was diagonal, the improvement was considerable for both the corrected and CoCo lasso. In contrast, when $\bm{\Sigma}_u$ was block-diagonal, the most remarkable improvement was made for the CoCo lasso only.  In all the settings, the PMSc$_{\text{FS}}$-CoCo estimators had the smallest $\ell_2$ estimation error. Finally, regarding computation time, it was not surprising that the SISc was still computationally the fastest screening method. The two-stage PMSc$_{\text{FS}}$-based estimators were also fast to compute, and when $p$ increased to $2000$, the increase in computation time for PMSc and SISc was small compared to that of the one-stage estimators.
\begin{table}[ht]
	\centering
	\caption{Performance of the one-stage and two-stage estimators in the simulation study based on false positive rate (FPR, in percentage), false negative rate (FNR, in percentage) $\ell_2$ error, and computation time (in seconds) when $\bm{\Sigma}_x$ has an homogeneous structure with $\rho_x = 0.5$. The CoCo estimator was only computed when the number of variables was no more than $1000$ (either in 1st or 2nd step). }
	\medskip
	\resizebox{0.85\textwidth}{!}{\begin{tabular}{lll rr rrr r}
			\toprule[1.5pt]
			$\bm\Sigma_u$  & $p$ & Estimator &  \multicolumn{2}{c}{1st step} &
			\multicolumn{3}{c}{2nd step} & Time  \\
			& & & FPR & FNR & FPR & FNR & $\ell_2$& \\
			\cmidrule(lr){4-5} \cmidrule(lr){6-8} 
		
		Diagonal & 1000 & One-stage Corrected &   - &   - &  20.0 &   0.1 & 0.96 &  18.9 \\ 
		&  & PMSc$_{\text{CV}}$-Corrected & 100.0 &   0.0 &  19.7 &   0.2 & 0.95 &  31.4 \\ 
		&  & PMSc$_{\text{FS}}$-Corrected &  7.6 &   3.2 &   1.6 &   3.2 & 0.56 &   6.8 \\ 
		&  & SISc-Corrected &   7.6 &   6.7 &   1.0 &   6.7 & 0.54 &   2.2 \\ 
		&  & One-stage CoCo &   - &   - &  11.5 &   0.0 & 0.77 & 481.5 \\ 
		&  & PMSc$_{\text{CV}}$-CoCo & 100.0 &   0.0 &  11.5 &   0.0 & 0.77 & 489.4 \\ 
		&  & PMSc$_{\text{FS}}$-CoCo &   7.6 &   3.2 &   1.0 &   3.2 & 0.45 &   7.2 \\ 
		&  & SISc-CoCo &   7.6 &   6.7 &   1.0 &   6.7 & 0.52 &   2.9 \\ 
		& 2000 & One-stage Corrected &   - &   - &  29.1 &   1.1 & 1.57 & 49.9 \\ 
		&  & PMSc$_{\text{CV}}$-Corrected & 100.0 &   0.0 &  28.3 &   1.0 & 1.53 & 74.5 \\ 
		&  & PMSc$_{\text{FS}}$-Corrected &   3.8 &  4.0 &   0.7 &  4.0 & 0.62 &  9.6 \\ 
		&  & SISc-Corrected &   3.8 &  11.0 &   0.5 &  11.0 & 0.61 &  3.6 \\ 
		&  & PMSc$_{\text{CV}}$-CoCo & 100.0 &   0.0 &   - &   - &  - &  - \\ 
		&  & PMSc$_{\text{FS}}$-CoCo &   3.8 &  4.0 &   0.5 & 4.0 & 0.45 &  10.2 \\ 
		&  & SISc-CoCo &   3.8 &  11.0 &   0.5 &  11.0 & 0.60 &  4.6 \\ \\
		
		\multirow{2}{1.2cm}{Block diagonal} & 1000 & One-stage Corrected &   - &   - &  23.5 &   1.1 & 1.25 &  22.7 \\ 
		&  & PMSc$_{\text{CV}}$-Corrected & 100.0 &   0.0 &  25.6 &   0.8 & 1.30 &  37.0 \\ 
		&  & PMSc$_{\text{FS}}$-Corrected &   7.6 &  3.4 &   3.0 &  3.6 & 1.15 &   6.7 \\ 
		&  & SISc-Corrected &   7.6 &  14.2 &   1.5 &  14.2 & 0.97 &   2.4 \\ 
		&  & One-stage CoCo &   - &   - &  14.3 &   0.4 & 1.23 & 417.2 \\ 
		&  & PMSc$_{\text{CV}}$-CoCo & 100.0 &   0.0 &  14.3 &   0.5 & 1.23 & 427.9 \\ 
		&  & PMSc$_{\text{FS}}$-CoCo &   7.6 &  3.4 &   0.9 &  3.6 & 0.68 &   8.2 \\ 
		&  & SISc-CoCo &   7.6 &  14.2 &   0.9 &  14.2 & 0.84 &   3.6 \\[3pt]
		 
		& 2000 & One-stage Corrected &   - &   - &  32.4 &   3.0 & 1.86 & 52.5 \\ 
		&  & PMSc$_{\text{CV}}$-Corrected & 100.0 &   0.0 &  31.8 &   3.3 & 1.84 & 79.8 \\ 
		&  & PMSc$_{\text{FS}}$-Corrected &   3.8 &  6.6 &   1.6 &  6.6 & 1.21 & 9.4  \\ 
		&  & SISc-Corrected &   3.8 &  21.4 &   0.8 &  21.4 & 1.09 &  3.8 \\ 
		&  & PMSc$_{\text{CV}}$-CoCo & 100.0 &   0.0 &   - &   - &  - &  - \\ 
		&  & PMSc$_{\text{FS}}$-CoCo &  3.8 &  6.6 &   0.5 &  6.7 & 0.73 &  10.8 \\ 
		&  & SISc-CoCo &   3.8 &  21.4 &   0.5 &  21.4 & 0.97 &  5.3 \\ 
		\bottomrule[1.5pt]
	\end{tabular}}
\label{tab:HomogenSigmax_0.5_short}
\end{table}

In summary, the simulation study both confirms the theoretical results for the screening procedures when the true covariates are not highly correlated, and demonstrates the superior finite sample performance, computational efficiency and scalability of the proposed two-stage estimators compared to the one-stage estimators. Among the three screening methods implemented in the simulations, the PMSc$_{\text{FS}}$ was the most reliable and efficient screening method, taking account of both performance metrics and computation time when the true covariates were either moderately or highly correlated.  
\FloatBarrier
\section{Analysis of microarray data}
\label{section:application}
We applied the proposed methodology to analyze the motivating Bone Mineral Density Data consisting of gene expression measurements of $54,675$ probe sets and bone mineral density (BMD) for $n=84$ Norwegian women. The dataset is publicly available at the European Bioinformatics Institute ArrayExpress repository under the access number E-MEXP-1618. Microarray measurements are known to be noisy \citep{rocke2001model}; although biological variation in the data is usually of primary interest to investigators, it can be obscured by measurement errors coming from many sources \citep{zakharkin2005sources}. Furthermore, a distinctive feature of the Affymetrix microarray dataset is that multiple probes are used to measure each gene expression; hence, these replicated measurements can be used to estimate the covariance matrix of measurement errors $\bm\Sigma_u$. 

We followed the procedures described in \citet{sorensen2015measurement}, \citet{nghiem2019simselex} and \citet{romeo2019model} to process the raw data using the BGX package of \citet{hein2005bgx}, and assumed the measurement error on each gene was mutually independent from that on the other. As a result, the measurement error covariance matrix $\bm{\Sigma}_u$ was set to be diagonal. Similar to \citet{sorensen2015measurement}, we also kept $p=993$ genes that had between-patient variability greater than measurement error variance, i.e the noise-to-signal ratio was smaller than $1$. After processing, we obtained the surrogate matrix $\bW$ along with $\bm{\Sigma}_u$. More details on the raw data processing and the estimated covariance matrix of measurement error can be found in Appendix B. The response variable was chosen to be the (centered) total hip T-score. 
 
We computed the eight estimators as in the simulation study. For the screening step, the PMSc$_{\text{CV}}$ procedure selected $610$ genes, so it did not reduce the number of dimensions by any great amount. This was likely due to the high correlation among the gene expressions, similar to the case when $\bm\Sigma_x$ had a homogeneous structure in the simulation. Table \ref{tab: data analysis} shows the number of genes selected by one-stage and two-stage estimators, as well as the $\ell_2$ norm of the estimated coefficients. 
\begin{table}[t]
	\centering
	\caption{The number of selected genes and the corresponding $\ell_2$ norm of the estimated coefficients obtained from each estimator in the microarray data analysis. Computation time is based on implementation on a laptop with one Dual-Core Intel Core i5 2.7GHz processor and 8GB RAM}
	\label{tab: data analysis}
	\medskip 
\begin{tabular}{lccccc}
	\toprule[1.5pt]
	Estimator &  \multicolumn{2}{c}{\# of selected genes} & $\ell_2$ norm  & \multicolumn{2}{c}{Time (in seconds)} \\
	& 1st step & 2nd step  & & 1st step & 2nd step   \\
	\hline
	One-stage Corrected & - & 108  & 0.33 & 0.00 & 85.97 \\ 
	PMSc$_{\text{CV}}$-Corrected & 798 & 104  &  0.34 & 74.02 &  43.48 \\
	PMSc$_{\text{FS}}$-Corrected & 19 &  4  &  2.10 & 6.54 & 2.49 \\ 
	SISc-Corrected & 19 & 4  & 2.00 & 0.73  & 2.49 \\[4pt]
	
	One-stage CoCo & - & 18  &   0.76 & 0.00 & 2702.81  \\ 
	PMSc$_{\text{CV}}$-CoCo & 798 & 17 & 0.78 & 74.02 & 1648.86 \\ 
	PMSc$_{\text{FS}}$-CoCo & 19 & 6  &  1.05 & 6.54 & 0.21   \\ 
	SISc-CoCo & 19 & 5  &  1.12 & 0.73 & 0.30  \\ 
	\bottomrule[1.5pt]
\end{tabular} 
\end{table}
It can be seen that while the one-stage corrected lasso and PMSc$_{\text{CV}}$-Corrected estimators select many more genes than all the other estimators, the $\ell_2$ norms of the corresponding estimated coefficients were smaller or equivalent to that of the other estimators; hence, the one-stage corrected lasso and PMSc$_{\text{CV}}$-Corrected estimators likely contained many false positives. Among the $p=993$ genes, there were 15 genes selected by one estimator, 113 genes selected by two estimators, 1 gene selected by 3 estimators, 1 gene selected by 4 estimators, and notably 3 genes selected by 6 estimators. Table \ref{tab: rank_est} demonstrates that the estimated coefficients associated with these three genes had larger magnitude than those of other selected genes. Furthermore, compared to the one-stage CoCo lasso and PMSc$_{\text{CV}}$-CoCo estimators, the two-stage  PMSc$_{\text{FS}}$-CoCo and SISc-CoCo estimators magnified the effects of these three genes. These estimators might contain some false negatives, but they help practitioners obtain stronger signals of the important variables that are present, as has been noted in the literature of measurement error correction in high dimensional settings; see \citet{sorensen2015measurement} and \citet{nghiem2019simselex}. Finally, regarding computation time, the two-stage estimators were generally faster to compute, with the exception of the PMSc$_{\text{CV}}$-based estimators; the gain in computation time was the largest when the CoCo estimator was used in the second stage.  

\begin{table}[t]
	\centering 
	\caption{The estimated coefficients and their corresponding rank in terms of magnitude for the three genes most frequently selected by most estimators.}
	\label{tab: rank_est}
	\medskip
	\begin{tabular}{l ccc ccc}
		\toprule[1.5pt]
		Estimator & \multicolumn{3}{c}{Estimate} & \multicolumn{3}{c}{Rank (1 = largest) }	\\
		& Gene 1 & Gene 2 & Gene 3 & Gene 1 & Gene 2 & Gene 3\\
		\cmidrule{2-7}
		PMSc$_{\text{FS}}$-Corrected &  0.96 & -1.54 & 0.62 & 2 & 1 & 4 \\
		SISc-Corrected & 0.86 & -1.43 & 0.63 & 2 & 1 & 4 \\
		One-stage CoCo & 0.61 & -0.25 & 0.27 & 1 & 3 & 2  \\ 
		PMSc$_{\text{CV}}$-CoCo & 0.62 & -0.28 & 0.29 & 1 & 3 & 2   \\ 
		PMSc$_{\text{FS}}$-CoCo &  0.78 & -0.47 & 0.29 & 1 & 2 & 4  \\ 
		SISc-CoCo & 0.49 & -0.94 & 0.18 & 2 & 1 & 4   \\ 
		\bottomrule[1.5pt]
		\end{tabular}
\end{table}
\FloatBarrier
\section{Conclusion}
\label{section:conclusion}
This paper proposes two screening procedures for linear errors-in-variables models in high dimensional settings, namely the corrected penalized marginal regression and the corrected sure independence screening procedures. Both procedures are based on fitting corrected marginal regressions of the outcome on each contaminated covariate, which could be computed efficiently in high dimensions. Under mild technical conditions, these procedures are shown to achieve screening consistency, meaning all the important variables are fully retained. Under a stronger condition of partial orthogonality  among the true covariates, we further illustrate that the corrected penalized marginal screening approach (using the bridge penalty) can achieve full selection consistency. We demonstrated the advantages of these screening procedures in practice through a simulation study and an analysis of a microarray data concerning gene expressions of bone mineral density for Norweigian women, both of which showed that the proposed screening procedures reduced computation time considerably and/or improved performance metrics for both estimation and variable selection. 

Future research could aim at reducing the false negative rates for screening procedures when the true covariates are highly collinear; for such a setting, an iterative corrected marginal screening procedure similar to the iterative sure independent screening of \citet{fan2008sure} may be considered. Furthermore, the screening procedures presented in this paper can be extended to more complicated errors-in-variables models, such as generalized linear models and non-parametric regression settings, although it may be more challenging to correct for measurement errors in marginal regressions of the outcome on each contaminated covariate in these models.





%
\bibliographystyle{apalike} 
\bibliography{citation}

\begin{thebibliography}{}

\bibitem[Barut et~al., 2016]{barut2016conditional}
Barut, E., Fan, J., and Verhasselt, A. (2016).
\newblock Conditional sure independence screening.
\newblock {\em Journal of the American Statistical Association},
  111(515):1266--1277.

\bibitem[Belloni et~al., 2017]{belloni2017linear}
Belloni, A., Rosenbaum, M., and Tsybakov, A.~B. (2017).
\newblock Linear and conic programming estimators in high dimensional
  errors-in-variables models.
\newblock {\em Journal of the Royal Statistical Society: Series B (Statistical
  Methodology)}, 79(3):939--956.

\bibitem[Brown et~al., 2019]{brown2019meboost}
Brown, B., Weaver, T., and Wolfson, J. (2019).
\newblock Meboost: Variable selection in the presence of measurement error.
\newblock {\em Statistics in Medicine}, 38(15):2705--2718.

\bibitem[Byrd and McGee, 2019]{byrd2019simple}
Byrd, M. and McGee, M. (2019).
\newblock A simple correction procedure for high-dimensional general linear
  models with measurement error.
\newblock {\em arXiv preprint arXiv:1912.11740}.

\bibitem[Carroll et~al., 2006]{carroll2006measurement}
Carroll, R.~J., Ruppert, D., Stefanski, L.~A., and Crainiceanu, C.~M. (2006).
\newblock {\em Measurement Error in Nonlinear Models: A Modern Perspective}.
\newblock CRC press.

\bibitem[Cui et~al., 2015]{cui2015model}
Cui, H., Li, R., and Zhong, W. (2015).
\newblock Model-free feature screening for ultrahigh dimensional discriminant
  analysis.
\newblock {\em Journal of the American Statistical Association},
  110(510):630--641.

\bibitem[Datta and Zou, 2020]{datta2020note}
Datta, A. and Zou, H. (2020).
\newblock A note on cross-validation for lasso under measurement errors.
\newblock {\em Technometrics}, 62(4):549--556.

\bibitem[Datta et~al., 2017]{datta2017cocolasso}
Datta, A., Zou, H., et~al. (2017).
\newblock Cocolasso for high-dimensional error-in-variables regression.
\newblock {\em The Annals of Statistics}, 45(6):2400--2426.

\bibitem[Do et~al., 2011]{do2011web}
Do, C.~B., Tung, J.~Y., Dorfman, E., Kiefer, A.~K., Drabant, E.~M., Francke,
  U., Mountain, J.~L., Goldman, S.~M., Tanner, C.~M., Langston, J.~W., et~al.
  (2011).
\newblock Web-based genome-wide association study identifies two novel loci and
  a substantial genetic component for parkinson's disease.
\newblock {\em PLoS Genet}, 7(6):e1002141.

\bibitem[Fan and Lv, 2008]{fan2008sure}
Fan, J. and Lv, J. (2008).
\newblock Sure independence screening for ultrahigh dimensional feature space.
\newblock {\em Journal of the Royal Statistical Society: Series B (Statistical
  Methodology)}, 70(5):849--911.

\bibitem[Fan et~al., 2004]{fan2004nonconcave}
Fan, J., Peng, H., et~al. (2004).
\newblock Nonconcave penalized likelihood with a diverging number of
  parameters.
\newblock {\em The Annals of Statistics}, 32(3):928--961.

\bibitem[Fan et~al., 2010]{fan2010sure}
Fan, J., Song, R., et~al. (2010).
\newblock Sure independence screening in generalized linear models with
  np-dimensionality.
\newblock {\em The Annals of Statistics}, 38(6):3567--3604.

\bibitem[Frank and Friedman, 1993]{frank1993statistical}
Frank, L.~E. and Friedman, J.~H. (1993).
\newblock A statistical view of some chemometrics regression tools.
\newblock {\em Technometrics}, 35(2):109--135.

\bibitem[Hein et~al., 2005]{hein2005bgx}
Hein, A.-M.~K., Richardson, S., Causton, H.~C., Ambler, G.~K., and Green, P.~J.
  (2005).
\newblock Bgx: {A} fully {B}ayesian integrated approach to the analysis of
  {A}ffymetrix {G}enechip data.
\newblock {\em Biostatistics}, 6(3):349--373.

\bibitem[Huang et~al., 2008]{huang2008asymptotic}
Huang, J., Horowitz, J.~L., Ma, S., et~al. (2008).
\newblock Asymptotic properties of bridge estimators in sparse high-dimensional
  regression models.
\newblock {\em The Annals of Statistics}, 36(2):587--613.

\bibitem[Huang et~al., 2009]{huang2009group}
Huang, J., Ma, S., Xie, H., and Zhang, C.-H. (2009).
\newblock A group bridge approach for variable selection.
\newblock {\em Biometrika}, 96(2):339--355.

\bibitem[Ida et~al., 2019]{ida2019fast}
Ida, Y., Fujiwara, Y., and Kashima, H. (2019).
\newblock Fast sparse group lasso.
\newblock In Wallach, H., Larochelle, H., Beygelzimer, A., d\textquotesingle
  Alch\'{e}-Buc, F., Fox, E., and Garnett, R., editors, {\em Advances in Neural
  Information Processing Systems}, volume~32, pages 1702--1710.

\bibitem[Kaul et~al., 2016]{kaul2016two}
Kaul, A., Koul, H.~L., Chawla, A., and Lahiri, S.~N. (2016).
\newblock Two stage non-penalized corrected least squares for high dimensional
  linear models with measurement error or missing covariates.
\newblock {\em arXiv preprint arXiv:1605.03154}.

\bibitem[Li et~al., 2012]{li2012robust}
Li, G., Peng, H., Zhang, J., Zhu, L., et~al. (2012).
\newblock Robust rank correlation based screening.
\newblock {\em The Annals of Statistics}, 40(3):1846--1877.

\bibitem[Loh and Wainwright, 2012]{loh2012}
Loh, P.-L. and Wainwright, M.~J. (2012).
\newblock High-dimensional regression with noisy and missing data: Provable
  guarantees with nonconvexity.
\newblock {\em The Annals of Statistics}, 40(3):1637--1664.

\bibitem[Nghiem and Potgieter, 2019]{nghiem2019simselex}
Nghiem, L. and Potgieter, C. (2019).
\newblock Simulation-selection-extrapolation: Estimation in high-dimensional
  errors-in-variables models.
\newblock {\em Biometrics}, 75(0):1133--1144.

\bibitem[Piironen et~al., 2017]{piironen2017sparsity}
Piironen, J., Vehtari, A., et~al. (2017).
\newblock Sparsity information and regularization in the horseshoe and other
  shrinkage priors.
\newblock {\em Electronic Journal of Statistics}, 11(2):5018--5051.

\bibitem[Polson et~al., 2014]{polson2014bayesian}
Polson, N.~G., Scott, J.~G., and Windle, J. (2014).
\newblock The {B}ayesian bridge.
\newblock {\em Journal of the Royal Statistical Society: Series B (Statistical
  Methodology)}, 76:713--733.

\bibitem[Rocke and Durbin, 2001]{rocke2001model}
Rocke, D.~M. and Durbin, B. (2001).
\newblock A model for measurement error for gene expression arrays.
\newblock {\em Journal of Computational Biology}, 8(6):557--569.

\bibitem[Romeo and Thoresen, 2019]{romeo2019model}
Romeo, G. and Thoresen, M. (2019).
\newblock Model selection in high-dimensional noisy data: a simulation study.
\newblock {\em Journal of Statistical Computation and Simulation},
  89(11):2031--2050.

\bibitem[Rosenbaum et~al., 2010]{rosenbaum2010sparse}
Rosenbaum, M., Tsybakov, A.~B., et~al. (2010).
\newblock Sparse recovery under matrix uncertainty.
\newblock {\em The Annals of Statistics}, 38(5):2620--2651.

\bibitem[Rosenbaum et~al., 2013]{rosenbaum2013improved}
Rosenbaum, M., Tsybakov, A.~B., et~al. (2013).
\newblock Improved {m}atrix {u}ncertainty selector.
\newblock In {\em From Probability to Statistics and Back: High-Dimensional
  Models and Processes--A Festschrift in Honor of Jon A. Wellner}, pages
  276--290. Institute of Mathematical Statistics.

\bibitem[Simon et~al., 2013]{simon2013sparse}
Simon, N., Friedman, J., Hastie, T., and Tibshirani, R. (2013).
\newblock A sparse-group lasso.
\newblock {\em Journal of Computational and Graphical Statistics},
  22(2):231--245.

\bibitem[S{\o}rensen et~al., 2015]{sorensen2015measurement}
S{\o}rensen, {\O}., Frigessi, A., and Thoresen, M. (2015).
\newblock Measurement error in {L}asso: Impact and likelihood bias correction.
\newblock {\em Statistica Sinica}, 25:809--829.

\bibitem[Tibshirani, 1996]{tibshirani1996regression}
Tibshirani, R. (1996).
\newblock Regression shrinkage and selection via the lasso.
\newblock {\em Journal of the Royal Statistical Society: Series B (Statistical
  Methodology)}, 58(1):267--288.

\bibitem[Wainwright, 2019]{wainwright2019high}
Wainwright, M.~J. (2019).
\newblock {\em High-dimensional Statistics: A Non-Asymptotic Viewpoint},
  volume~48.
\newblock Cambridge University Press.

\bibitem[Wen et~al., 2018]{wen2018sure}
Wen, C., Pan, W., Huang, M., and Wang, X. (2018).
\newblock Sure independence screening adjusted for confounding covariates with
  ultrahigh dimensional data.
\newblock {\em Statistica Sinica}, 28:293--317.

\bibitem[Zakharkin et~al., 2005]{zakharkin2005sources}
Zakharkin, S.~O., Kim, K., Mehta, T., Chen, L., Barnes, S., Scheirer, K.~E.,
  Parrish, R.~S., Allison, D.~B., and Page, G.~P. (2005).
\newblock Sources of variation in {A}ffymetrix microarray experiments.
\newblock {\em BMC bioinformatics}, 6(1):1--11.

\bibitem[Zhou et~al., 2018]{zhou2018brain}
Zhou, T., Thung, K.-H., Liu, M., and Shen, D. (2018).
\newblock Brain-wide genome-wide association study for alzheimer's disease via
  joint projection learning and sparse regression model.
\newblock {\em IEEE Transactions on Biomedical Engineering}, 66(1):165--175.

\bibitem[Zhu et~al., 2011]{zhu2011model}
Zhu, L.-P., Li, L., Li, R., and Zhu, L.-X. (2011).
\newblock Model-free feature screening for ultrahigh-dimensional data.
\newblock {\em Journal of the American Statistical Association},
  106(496):1464--1475.

\end{thebibliography}

\newpage
\begin{center}
\textbf{\Large Appendix}	
\end{center}
\setcounter{section}{0}
\renewcommand\thesection{\Alph{section}}
\renewcommand\thetable{\thesection.\arabic{table}}
\counterwithin{table}{section}
    
\section{Technical proofs}
We first recall the technical conditions that are given in the main paper. 
\begin{enumerate}[label=(C\arabic*), ref = C\theenumi]	   
	
	\item  \label{cond:epsilonsubGaussian} The model error terms $\varepsilon_{1}, \varepsilon_{2}, \ldots \varepsilon_n$ are iid sub-Gaussian random variables with mean zero and finite variance $\sigma^{2}$, i.e there exists a finite constant $\sigma_*>0$  such that for all $t\in \mathbb{R}$,
	$$ \mathbb{E}\left\{\exp\left(t\varepsilon_i\right)\right\} \leq \exp \left(\frac{\sigma_*^2t^2}{2}\right),~ i = 1,\ldots, n,$$ 
	where the constant $\sigma_*^2$ is referred to as the ``variance proxy`` for $\varepsilon_i$. 
	\label{cond:B1}
	
	\item \label{cond: MEsubGaussian} The measurement errors $U_{ij},~i=1,\ldots,n$ are independent sub-Gaussian random variables with variance proxy $\sigma_*^2$, $j = 1,\ldots, p$. 
	\label{cond:A2} 
	
	\item  \label{cond:B4} \label{cond:boundonbetak} There exist constants $b_0$ and $b_1$ such that $\displaystyle 0 < b_0 \leq \min _{k \in \mathcal{S}}\left|\beta_{0 k}\right| \leq \max _{k \in \mathcal{S}}\left|\beta_{0 k}\right| \leq b_{1} \leq \infty. $ 
	
	\item \label{cond:B6} There exist constants $C_1$ and $C_2$ such that 
	$\displaystyle 0 < C_1 \leq \min_{j} n^{-1} \sum_{i=1}^{n} X_{ij}^2 < \max_{j} n^{-1}\sum_{i=1}^{n} X_{ij}^2 \leq C_2 < \infty. \label{cond:boundonxsquare} 
	$
	
	\item \label{cond:new} There exists a constant $\xi_0$ such that $\displaystyle\min _{k \in \mathcal{S}}|\tilde{\xi}_{n j}|>\xi_{0}>0 $ for all $n$. \label{cond:boundonxandy}
	
	\item There exists constants $d_{0}>0 $ and $0\leq\theta\leq 1/2$ such that
	$$
	\left|n^{-1} \sum_{i=1}^{n} X_{i j} X_{i k}\right| \leq d_{0} n^{-\theta}, \quad j \in \mathcal{S}^c,~ k \in \mathcal{S},
	$$ 
	\label{cond:B2}
	\item Assume the tuning parameter $\lambda_n$ for PMSc satisfies:
	\begin{enumerate}[label = (\alph*), ref=\theenumi\alph*]
		\item $\lambda_n \to 0$ and  $\lambda_{n} n^{\theta(2-\alpha) } s^{\alpha-2} \rightarrow \infty$, \label{cond:C9a}
		\item  $\log \left(2m\right)= o(1) \times \left\{\lambda_{n} n^{(2-\alpha) / 2}\right\}^{1 /(2-\alpha)},$ where $ m = p-s$. \label{cond:C9b}
	\end{enumerate}
	
	\item $p >n$, $\log(p)/n \to C_1$ with $C_1$ being a constant as both $n$ and $p$ diverge to $\infty$.	\label{cond:np}
	
	
	\item  Let $\mathbf{Z} = \bU \bm\Sigma_u^{-1/2}$. Then 
	\begin{enumerate}[label = (\alph*), ref=\theenumi\alph*]
		\item \label{cond:MEconcentration} For some  constants $c_1, c_2 >1$ and $D> 0$, the matrix  $\mathbf{Z}$ follows a spherical distribution satisfying
		\begin{equation}
			\mathbb{P}\left(\lambda_{\max }\left(\tilde{p}^{-1} \tilde{\mathbf{Z}} \widetilde{\mathbf{Z}}^{\top}\right)>c_{2} \text { and } \lambda_{\min }\left(\widetilde{p}^{-1} \tilde{\mathbf{Z}} \tilde{\mathbf{Z}}^{\top}\right)<1 / c_{2}\right) \leq e^{-Dn},
			\label{cond: concentrationproperty}
		\end{equation}	 
		for any $ n\times \tilde{p}$ submatrix $\tilde{\mathbf{Z}}$ of $\mathbf{Z}$ with $c_1n \leq \tilde{p} \leq p$.
		\item \label{cond:MEeigenvalue} There exist some constants $\tau_1>0$ and $c_3>0$, such that 
		$
		\lambda_{\max }(\bm{\Sigma}_u) \leq c_3n^{\tau_1}.
		$
	\end{enumerate}
	\item \label{cond:boundoneigX} There exists positive constants $c_4>0$ and  $\tau_2 > 0$, such that
	$
	\lambda_\text{max}\left(n^{-1}\bX^\top \bX\right) \leq c_4n^{\tau_2}.
	$
	\item \label{cond:tau} $\tau_1 + \tau_2 + \log_n(s)  <1$, where  $\log_n(s)$ denotes the logarithm base $n$ of $s$. 
\end{enumerate}

\subsection{Proof of Theorem 1}

First, we state the  following lemma from Huang et al. (2008) that will be used in the proof. 
\begin{lemma}
	Let $g(u) = u^2 - 2au + \lambda \vert u \vert ^\alpha $ where $a\neq 0$, $\lambda \geq 0 $ and $0 < \alpha < 1$. Define
	$$c_{\alpha}=\left(\frac{2}{2-\alpha}\right)\left\{\frac{2(1-\alpha)}{2-\alpha}\right\}^{1-\alpha}.$$
	Then $\arg\min(g) =0 $ if and only if $\lambda>c_{\alpha}|a|^{2-\alpha}$.
	\label{lemma1}
\end{lemma}

Additionally, we state the following properties of sub-Gaussian random variables from \citet{wainwright2019high}:
\begin{enumerate}[label=(B\arabic*)]
	\item Let $S_1, \ldots, S_n$  be zero-mean independent sub-Gaussian random variables with variance proxy  $\sigma_0^2$. Denote $\bm{v} = (v_1, \ldots, v_n) \in {R}^{n}$. Then
	$\sum_{i=1}^{n}v_iS_i$ is sub-Gaussian with variance bounded by $\left(\sum_{i=1}^{n}v_i^2\right)\sigma_0^2$. \label{prop:B2}
	\item \label{prop:B3} Let $S_1, \ldots, S_n$ be zero-mean sub-Gaussian random variables (not necessarily independent) with common variance proxy $\sigma_0^2$. Then $$\mathbb{E}\left(\max_{i=1,\ldots, n} \vert S_i \vert\right) \leq \sigma_0 \sqrt{2\log(2n)}.$$ 
	\item If $S_1$ and $S_2$ are zero-mean independent sub-Gaussian random variables with finite variance proxies $\sigma_{01}^2$ and $\sigma_{02}^2$ respectively, then $V=S_1 S_2$ is sub-exponential with (finite) parameter $(\nu_0, t_0)$, meaning that
	$\mathbb{E}(\exp(sV)) \leq \exp(\frac{1}{2}\nu_0^2s^2)$ for all $\vert s\vert > t_0.$ \label{prop:B4} 
	\item Let $S_1, \ldots, S_n$ be sub-exponential random variables (not necessarily independent) with common  parameter $(\nu_0, t_0)$. Then $$\mathbb{E}\left(\max_{i=1,\ldots, n} \vert S_i \vert\right) \leq \nu_0 \sqrt{2\log(2n)} + t_0 \log(2n).$$ \label{prop:B5}
\end{enumerate}

\begin{proof}
	Recall that,
	$$
	\tilde{\xi}_{n j}=n^{-1} \sum_{i=1}^{n}\left(\sum_{k=1}^{s} X_{ik}\boldsymbol{\beta}_{0k}\right) X_{i j}
	, \quad \xi_{n j}=\tilde{\xi}_{n j} + n^{-1}\sum_{i=1}^{n}\left(\sum_{k=1}^{s} X_{ik}\boldsymbol{\beta}_{0k}\right) U_{i j},
	$$
	so we have
	\begin{align*} 
		L(\boldsymbol{\beta}) &=\dfrac{1}{n}\sum_{j=1}^{p} \sum_{i=1}^{n}\left(y_{i}-W_{i j} \beta_{j}\right)^{2}-\sum_{j=1}^{p}\sigma_j^2\beta_j^2 +\lambda_{n} \sum_{j=1}^{p}\left|\beta_{j}\right|^{\alpha} \\
		& =\sum_{j=1}^{p} \sum_{i=1}^{n}\dfrac{1}{n}\left(\sum_{k=1}^{s}X_{ik}\beta_{0k}+\varepsilon_i-W_{i j} \beta_{j}\right)^{2}-\sum_{j=1}^{p}\sigma_j^2\beta_j^2 +\lambda_{n} \sum_{j=1}^{p}\left|\beta_{j}\right|^{\alpha} \\ &=\sum_{j=1}^{p}\left[\dfrac{1}{n}\sum_{i=1}^{n} \varepsilon_{i}^{2}+\left(\dfrac{1}{n}\sum_{i=1}^{n} W_{ij}^2 - \sigma_j^2 \right) \beta_{j}^{2}-\frac{2}{n}\left(\bm\varepsilon^{\top} \mathbf{a}_{j} + n \xi_{n j}\right) \beta_{j}+ \frac{1}{n} \sum_{i=1}^{n} \left(\sum_{k=1}^{s} X_{ik}\beta_{0k}\right)^2 + \lambda_{n}\left|\beta_{j}\right|^{\alpha}\right] 
	\end{align*}
	where $\textbf{a}_j = (W_{1j}, \ldots, W_{nj})^\top = (X_{1j}, \ldots, X_{nj})^\top + (U_{1j}, \ldots, U_{nj})^\top = \bX_j + \bU_j$. We can ignore the terms that do not contain $\bm\beta$, so minimizing $L(\bm{\beta})$ is equivalent to minimizing $\sum_{j=1}^{p}h_j(\beta_j)$, where
	$$
	h_j(\beta_j) = \left(\dfrac{1}{n}\sum_{i=1}^{n} W_{ij}^2 - \sigma_j^2 \right) \beta_{j}^{2}-\frac{2}{n}\left(\bm\varepsilon^{\top} \mathbf{a}_{j}+n \xi_{n j}\right) \beta_{j}+\lambda_{n}\left|\beta_{j}\right|^{\alpha}, \quad j=1,\ldots ,p.
	$$
	To simplify the notation, let $V_j = n^{-1}\sum_{i=1}^{n} W_{ij}^2 - \sigma_j^2$. By Lemma \ref{lemma1}, $\hat\beta_j=0$ is the only solution of $\arg\min_{\beta_j} h_j(\beta_j)$ if and only if
	\[
	V_j^{-1} \lambda_{n}>c_{\alpha}\left\{\left(nV_j\right)^{-1} \left|\bm\varepsilon^{\top} \mathbf{a}_{j}+n \xi_{n j}\right|\right\}^{2-\alpha}.
	\]
	By some algebra, letting $\phi_n = c_{\alpha}^{-1 /(2-\alpha)}\lambda_{n}^{1 /(2-\alpha)}n^{1/2}$, the above inequality is equivalent to 
	\[
	\phi_{n}>n^{-1 / 2}V_j^{(-1+\alpha)/(2-\alpha)} \left|\bm\varepsilon^{\top} \mathbf{a}_{j}+n \xi_{n j}\right|.
	\]
	Hence, Theorem 1 will follow if we can prove that 
	\begin{equation}
		\lim_{n \to \infty} \mathbb{P}\left\{\phi_{n}>n^{-1 / 2}\min_{j\in \mathcal{S}}V_j^{(-1+\alpha)/(2-\alpha)} \left|\bm\varepsilon^{\top} \mathbf{a}_{j}+n \xi_{n j}\right|\right\} \rightarrow 0.
		\label{eq:K1}
	\end{equation} 
	In order to prove \eqref{eq:K1}, note that as $ n\to \infty$, for every $j = 1,\ldots, p$, we have 
	$$ 
	\begin{aligned}
		\left\vert \left(\frac{1}{n} \sum_{i=1}^{n}W_{ij}^2 - \sigma_j^2\right) - \dfrac{1}{n}\sum_{i=1}^{n} X_{ij}^2 \right \vert & = \left \vert \frac{1}{n}\sum_{i=1}^{n} \left(X_{ij}+U_{ij}\right)^2 -  \sigma_j^2 - \dfrac{1}{n}\sum_{i=1}^{n} X_{ij}^2 \right\vert \\ &\leq \left\vert\dfrac{1}{n} \left(\sum_{j=1}^{n}U_{ij}^2\right) - \sigma_j^2\right\vert + \left\vert\dfrac{2}{n} \sum_{i=1}^{n} X_{ij} U_{ij}\right\vert = o_p(1),
	\end{aligned}
	$$
	by the law of large numbers and that $E(U_{ij}) = 0, E(U_{ij}^2) = \sigma_j^2$ and $E(U_{ij}^4) <\infty$. Let $\tau_j = (n^{-1} \sum_{i=1}^{n} X_{ij}^2)^{(-1+\alpha)/(2-\alpha)}$. By condition \eqref{cond:B6} and because $0 < \alpha < 1$, there exist constants $C_3$ and $C_4$ such that $C_3\leq \min_j\tau_j \leq \max_j \tau_j \leq C_4 < \infty$. It suffices to show that
	\begin{equation}
		\mathbb{P}\left\{\phi_{n}>n^{-1 / 2}\min_{j\in \mathcal{S}}\tau_j \left|\bm\varepsilon^{\top} \mathbf{a}_{j}+n \xi_{n j}\right|\right\} \rightarrow 0.
		\label{eq:K}
	\end{equation} 
	Now, to prove \eqref{eq:K},  we have
	\begin{equation}
		\begin{aligned}
			\mathbb{P}\left(\phi_{n}>\min_{j \in \mathcal{S}}\tau_j\left|n^{-1 / 2} \bm\varepsilon^{\top} \mathbf{a}_{j}+n^{1 / 2} \xi_{n j}\right|\right) & =\mathbb{P}\left(\bigcup_{j \in \mathcal{S}}\left\{\tau_j\left|n^{-1 / 2} \bm\varepsilon^{\top} \mathbf{a}_{j}+n^{1 / 2} \xi_{n j}\right|<\phi_{n}\right\}\right) \\ & \leq \sum_{j \in \mathcal{S}} \mathbb{P}\left(\tau_j\left|n^{-1 / 2} \bm\varepsilon^{\top} \mathbf{a}_{j}+n^{1 / 2} \xi_{n j}\right|<\phi_{n}\right) \\
			& \leq \sum_{j \in \mathcal{S}} \mathbb{P}\left(\left|n^{-1 / 2} \bm\varepsilon^{\top} \mathbf{a}_{j}+n^{1 / 2} \xi_{n j}\right|<C_3^{-1}\phi_{n}\right)
		\end{aligned}
		\label{eq8}
	\end{equation}
	where the last inequality follows from $\tau_j \geq C_3$. Then noting that for all $j \in \mathcal{S}$, we have
	\begin{equation}
		\mathbb{P}\left(\left|n^{-1 / 2} \bm\varepsilon^{\top} \mathbf{a}_{j}+n^{1 / 2} \xi_{n j}\right|<C_3^{-1}\phi_{n}\right) = 1-\mathbb{P}\left(\left|n^{-1 / 2} \bm\varepsilon^{\top} \mathbf{a}_{j}+n^{1 / 2} \xi_{n j}\right| \geq C_3^{-1}\phi_{n}\right),
		\label{eq:9}
	\end{equation}
	and hence 
	\begin{align*}
		& \mathbb{P}\left(\left|n^{-1 / 2} \bm\varepsilon^{\top} \mathbf{a}_{j}+n^{1 / 2} \xi_{n j}\right| \geq C_3^{-1}\phi_n\right) \geq \mathbb{P}\left(n^{1 / 2}\left|\xi_{n j}\right|-n^{-1 / 2}\left|\bm\varepsilon^{\top} \mathbf{a}_{j}\right| \geq C_3^{-1}\phi_n\right) \\ & =1-\mathbb{P}\left(n^{-1 / 2}\left|\bm\varepsilon^{\top} \mathbf{a}_{j}\right|>n^{1 / 2}\left|\xi_{n j}\right|-C_3^{-1}\phi_n\right) \\ & \geq 1- \mathbb{P}\left(n^{-1 / 2}\left|\bm\varepsilon^{\top} \mathbf{a}_{j}\right| > n^{1/2} \vert\tilde{\xi}_{nj}\vert - n^{-1/2} \left\vert \sum_{i=1}^{n} \left(\sum_{k=1}^{s} X_{ik}\beta_{0k}U_{ij}\right)\right\vert - C_3^{-1}\phi_n\right) \\ & \geq 1 - \mathbb{P}\left(n^{-1 / 2}\left|\bm\varepsilon^{\top} \mathbf{X}_{j}\right| + n^{-1/2} \left|\bm\varepsilon^{\top} \mathbf{U}_{j}\right| +  n^{-1/2} \left\vert \sum_{i=1}^{n} \left(\sum_{k=1}^{s} X_{ik}\beta_{0k}U_{ij}\right)\right\vert > n^{1/2} \vert \xi_0 \vert - C_3^{-1}\phi_n\right) \\ & 
		\geq 1 - \mathbb{P}\left(\left\vert n^{-1 / 2}\bm\varepsilon^{\top} \mathbf{X}_{j} + n^{-1/2}\bm\varepsilon^{\top} \mathbf{U}_{j} +  n^{-1/2}  \sum_{i=1}^{n} \left(\sum_{k=1}^{s} X_{ik}\beta_{0k}U_{ij}\right)\right\vert > n^{1/2}  \vert\xi_0 \vert - C_3^{-1}\phi_n\right) \\ & 
		\geq 1 - \frac{\mathbb{E}\left\{n^{-1 / 2}\bm\varepsilon^{\top} \mathbf{X}_{j} + n^{-1/2}\bm\varepsilon^{\top} \mathbf{U}_{j} +  n^{-1/2} b_1 \sum_{i=1}^{n} \left(\sum_{k=1}^{s} X_{ik}U_{ij}\right) 
			\right\}^2}{\left\{n^{1/2} \xi_0 - C_3^{-1}\phi_n\right\}^2} \\ &
		\geq 1 - \frac{ n^{-1}\text{Var}\left[\bm\varepsilon^{\top} \mathbf{X}_{j} + \bm\varepsilon^{\top} \mathbf{U}_{j} +   b_1 \sum_{i=1}^{n}  \left(\sum_{k=1}^{s} X_{ik}U_{ij}\right)\right]}{\left\{n^{1/2} \xi_0 - C_3^{-1}\phi_n\right\}^2}.
	\end{align*}
	Now, consider the numerator of the last expression. Note that, 
	$$
	\text{Cov}(\bm\varepsilon^{\top} \bX_j, \bm\varepsilon^{\top} \bU_j) = \sum_{i=1}^{n} X_{ij} \text{Cov}\left(\epsilon_i, \epsilon_i U_{ji}\right) = \sum_{i=1}^{n} X_{ij} \left[ E(\epsilon_i^2) E(U_{ij}) - E(\epsilon_i)E(\epsilon_iU_{ij})\right] = 0
	$$ and also
	$$
	\text{Cov}(X_{ik}U_{ij}, \bm\varepsilon^{\top} \bU_j) = \text{Cov}(X_{ik}U_{ij}, \varepsilon_i U_{ij}) = X_{ik} E(\varepsilon_i)E(U_{ij}^2) = 0.
	$$
	Hence, 
	$$
	\begin{aligned}
		\mathrm{Var}\left[\bm\varepsilon^{\top} \mathbf{X}_{j} + \bm\varepsilon^{\top} \mathbf{U}_{j} +   b_1 \sum_{i=1}^{n}  \left(\sum_{k=1}^{s} X_{ik}U_{ij}\right)\right] &= \textrm{Var}(\bm\varepsilon^{\top} \mathbf{X}_{j}) + \textrm{Var}(\bm\varepsilon^{\top} \mathbf{U}_{j}) + \textrm{Var}\left[b_1 \sum_{i=1}^{n}  \left(\sum_{k=1}^{s} X_{ik}U_{ij}\right)\right] \\ &= \sigma^2 \sum_{i=1}^{n} X_{ij}^2 + \sum_{i=1}^{n}E(\epsilon_i^2)E(U_{ij}^2) + b_1^2\sum_{i=1}^{n}\left(\sum_{k=1}^{s}X_{ik}\right)^2E(U_{ij}^2) \\& \leq nC_2\sigma^2 + n\sigma^2\sigma_*^2 + nb_1^2 C_2s\sigma_*^2 = nO(1) + nsO(1).
	\end{aligned}
	$$
	It follows that 
	$$
	\mathbb{P}\left(\left|n^{-1 / 2} \bm\varepsilon^{\top} \mathbf{a}_{j}+n^{1 / 2} \xi_{n j}\right|<C_3^{-1}\phi_{n}\right) \leq \dfrac{O(1) + s O(1)}{\left\{n^{1/2} \xi_0 - C_3^{-1}\phi_n\right\}^2}.
	$$ 
	Since $\lambda_n = o(1)$, we then have
	$$
	\frac{\phi_{n}}{n^{1/2}} =   O(1)\lambda_{n}^{1 /(2-\alpha)}= o(1),
	$$
	and  $n^{-1} s = o(1)$. Therefore, we have 
	$$\mathbb{P}\left(\phi_{n}>\min _{j \in \mathcal{S}}\left|n^{-1 / 2} \bm\varepsilon^{\top} \mathbf{a}_{j}+n^{1 / 2} \xi_{n j}\right|\right)=O(1) n^{-1} s^2  \rightarrow 0 $$
	and \eqref{eq:K} follows.
\end{proof}

\subsection{Proof of Theorem 2}

Using the same notation as in the proof of Theorem 1, it suffices to show that

\begin{equation}
	\mathbb{P}\left\{\phi_{n}>n^{-1 / 2}\max_{j\in \mathcal{S}^c}\tau_j \left|\bm\varepsilon^{\top} \mathbf{a}_{j}+n \xi_{n j}\right|\right\} \rightarrow 1.
	\label{eq:J}
\end{equation} 
To prove \eqref{eq:J}, by condition \eqref{cond:B2}, for all $j \in \mathcal{S}^c$, we obtain
\begin{equation*}
	n^{1 / 2}\left|\tilde\xi_{n j}\right| = n^{-1 / 2}\left|\sum_{k=1}^{s} \sum_{i=1}^{n} X_{i k} X_{i j}  \beta_{0 k}\right|  \leq n^{-1 / 2} b_{1} \left(\sum_{l=1}^{s}\left|\sum_{i=1}^{n} X_{i k} X_{i j}\right| \right)  \leq b_{1} d_{0} n^{1/2-\theta} s = c_1 n^{1/2-\theta} s
\end{equation*}
with $c_1 = b_1 d_0$. Next, 
$$
\begin{aligned}
	& \mathbb{P}\left(\phi_{n}>n^{-1 / 2} \max_{j \in \mathcal{S}^c}\tau_j\left|\bm\varepsilon^{\top} \mathbf{a}_{j}+n\xi_{n j}\right|\right)  \geq \mathbb{P}\left(C_4^{-1}\phi_{n}>n^{-1/ 2} \max _{j \in \mathcal{S}^c}\left|\bm\varepsilon^{\top}  \mathbf{a}_{j}\right|+n^{1 / 2} \max _{j \in \mathcal{S}^c}\left|\xi_{n j}\right|\right) 
	\\ 
	& \geq \mathbb{P}\left(C_4^{-1}\phi_{n}> n^{-1/2} \max _{j \in \mathcal{S}^c}\left|\bm\varepsilon^{\top} \mathbf{X}_{j} \right| +  n^{-1/2} \max_{j \in \mathcal{S}^c} \left|\bm\varepsilon^{\top} \bU_j \right|+ n^{1/2} \max_{j\in \mathcal{S}^c} \vert \tilde{\xi}_{nj} \vert  \right. \\ & ~\qquad \qquad \qquad \qquad \qquad \qquad \qquad \left. + \max_{j\in \mathcal{S}^c} \left\vert n^{-1/2} \sum_{i=1}^{n} \left(\sum_{k=1}^{s} X_{ik} \beta_{0k}\right) U_{ij} \right\vert \right)
	\\ 
	& \geq \mathbb{P}\left(C_4^{-1}\phi_{n}> n^{-1/2} \max _{j \in \mathcal{S}^c}\left|\bm\varepsilon^{\top} \mathbf{X}_{j} \right| +  n^{-1/2} \max_{j \in \mathcal{S}^c} \left|\bm\varepsilon^{\top} \bU_j \right|+ c_1 n^{1/2-\theta} s + b_1 \max_{j\in \mathcal{S}^c} \left\vert n^{-1/2} \sum_{i=1}^{n} \left(\sum_{k=1}^{s} X_{ik} \right) U_{ij} \right\vert \right)
	\\
	& = 1 - \\ & \mathbb{P}\left(n^{-1/2} \max _{j \in \mathcal{S}^c}\left|\bm\varepsilon^{\top} \mathbf{X}_{j} \right| +  n^{-1/2} \max_{j \in \mathcal{S}^c} \left|\bm\varepsilon^{\top} \bU_j \right| + b_1 \max_{j\in \mathcal{S}^c} \left\vert n^{-1/2} \sum_{i=1}^{n} \left(\sum_{k=1}^{s} X_{ik} \right) U_{ij} \right\vert \geq C_4^{-1}\phi_{n}-c_{1} n^{1/2-\theta} s\right)
	\\ & \geq 1-\frac{\mathbb{E}\left(n^{-1 / 2} \max _{j \in \mathcal{S}^c}\left|\bm\varepsilon^{\top} \bX_{j}\right| + n^{-1/2} \max_{j \in \mathcal{S}^c} \left|\bm\varepsilon^{\top} \bU_j \right| + b_1 \max_{j\in \mathcal{S}^c} \left\vert n^{-1/2} \sum_{i=1}^{n} \left(\sum_{k=1}^{s} X_{ik} \right) U_{ij} \right\vert\right)}{C_4^{-1}\phi_{n}-c_{1} n^{1/2-\theta} s}.
\end{aligned}
$$
Consider each term in the numerator. For the first term, by property \ref{prop:B2}, because $\varepsilon_i$ is sub-Gaussian with common variance  $\sigma^2,~ i=1,\ldots,n$, then $\bm\varepsilon^{\top} \bX_j$ is sub-Gaussian with variance $\sigma^2\sum_{i=1}^{n}X_{ij}^2$  and variance proxy $C_2n\sigma^2$. 
Therefore, by property \ref{prop:B4}, we have 
$$\mathbb{E}\left\{n^{-1 / 2} \max _{j \in \mathcal{S}^c}\left|\bm\varepsilon^{\top} \bX_{j}\right| \right\} \leq C_2^{1/2}\sigma \sqrt{\log(2m)} = O(1) \left\{\log(2m)\right\}^{1/2}.$$
For the second term, by property \ref{prop:B4}, each variable $\varepsilon_i {U}_{ij}$ is sub-exponential with parameter $(\nu_0, t_0)$, so $\bm\varepsilon^{\top} \bU_j = \sum_{i=1}^{n}\epsilon_i U_{ij}$ is sub-exponential with parameter $(n\nu_0,t_0)$. Hence, by property \ref{prop:B5}, we have
$$
\mathbb{E}\left\{n^{-1 / 2} \max _{j \in \mathcal{S}^c}\left|\bm\varepsilon^{\top} \bU_{j}\right| \right\} \leq \nu_0^{1/2} \sqrt{\log(2m)} + n^{-1/2} t_0 \log(2m).   
$$
For the third term, by a similar argument, each variable $\sum_{i=1}^{n} \left(\sum_{k=1}^{s} X_{ik} \right) U_{ij}$ is sub-Gaussian with variance proxy $C_2\sigma_*^2ns$. Hence, 
$$
\mathbb{E}\left\{\max_{j\in \mathcal{S}^c} \left\vert n^{-1/2} \sum_{i=1}^{n} \left(\sum_{k=1}^{s} X_{ik} \right) U_{ij} \right\vert\right\} \leq C_2^{1/2}\sigma_*\sqrt{s\log(2m)} = O(1) \sqrt{s\log(2m)}.
$$
Putting the results together, we have
\begin{equation}
	\mathbb{P}\left(\phi_{n}>n^{-1 / 2} \max_{j \in \mathcal{S}^c}\tau_j\left|\bm\varepsilon^{\top} \mathbf{a}_{j}+n\xi_{n j}\right|\right) \geq 1 - \frac{\log(2m)^{1/2} \left\{O(1) +\sqrt{s}+  n^{-1/2} \sqrt{\log(2m)}\right\}   
	}{{C_4^{-1}\phi_{n}-c_{1} n^{1/2-\theta} s}}.
\end{equation}
Next, by condition \eqref{cond:C9a}, 
\begin{equation*}
	\frac{n^{1/2-\theta}s}{C_4^{-1}\phi_{n}}=O(1) \left(\frac{s^{(2-\alpha)}}{\lambda_{n} n^{\theta(2-\alpha)}}\right)^{1 /(2-\alpha)}  = o(1), 
\end{equation*}
and by condition \eqref{cond:C9b}, 
$$
\frac{\log(2m)}{C_4^{-1}\phi_n} = O(1) \left(
\frac{\log(2m)^{(2-\alpha)}}{\lambda_{n} n^{(2-\alpha) / 2}}\right)^{1 /(2-\alpha)} = o (1).
$$
Therefore, $\sqrt{\log(2m)}/\left(C_4^{-1}\phi_n \right)= o(1)$ and also  $\sqrt{s\log(2m)}/\left(C_4^{-1}\phi_n\right) = o(1)$. Finally, we have 
\begin{equation*}
	\mathbb{P}\left(\phi_{n}>n^{-1 / 2} \max _{j \in \mathcal{S}^c}\tau_j\left|\bm\varepsilon^{\top} \mathbf{a}_{j}+n \xi_{n j}\right|\right)  \rightarrow 1,
\end{equation*}
as required. 
\subsection{Proof of Theorem 3}
Recall that the screening set for SISc is defined as
$$
\hat{\mathcal{Q}}_{\text{SISc}} = \left\{1 \leq i \leq p:\lvert\tilde\beta_{i}\rvert \text { is among the first }d = \floor{\gamma n} \text { largest of all }\right\}.
$$ In this proof, we will call this set $\hat{\mathcal{Q}}_\gamma$ to emphasize the dependence of the screening set on $\gamma$, while the subscript SISc is omitted to ease the notation. Following Fan \& Lv (2007), the proof consists of two main steps. In the first step,  we define the following set
$$
\widetilde{\mathcal{M}}_{\delta}^{1}=\left\{1 \leq i \leq p:\lvert\tilde\beta_{i}\rvert \text { is among the first }[\delta p] \text { largest of all }\right\},
$$
and prove that for some positive constant $C$, we have
\begin{equation}
	\mathbb{P}\left( \widetilde{\mathcal{M}}_{\delta}^{1} \supseteq \mathcal{S} \right) = O(p\exp(-Cn))
	\label{eq: main_SISc}.
\end{equation}
In the second step, we will apply the dimensional reduction procedure above successively until the number of covariates to be kept is smaller than $n$. 

\paragraph{Step 1.} We will prove \eqref{eq: main_SISc} by bounding $\Vert\tilde{\bbeta}\Vert_2^2$ from above and $\vert \tilde{\beta}_j \vert, ~ j \in \mathcal{S}$ from below. Note that 
\begin{equation}
	\tilde\beta_j = \dfrac{ n^{-1}\sum_{i=1}^{n}W_{ij}y_i}{n^{-1}\sum_{i=1}^{n} W_{ij}^2-\sigma_j^2}, \quad j =1,\ldots, p,
	\label{eq:tildebetaj}
\end{equation}
so for the denominator of \eqref{eq:tildebetaj}, we have
$$
\mathbb{E}\left(\dfrac{1}{n} \sum_{i=1}^{n}W_{ij}^2 - \sigma_j^2\right) = \dfrac{1}{n} \sum_{i=1}^{n} X_{ij}^2,
$$
and
$$
\dfrac{1}{n} \sum_{i=1}^{n}W_{ij}^2 - \sigma_j^2 - \dfrac{1}{n} \sum_{i=1}^{n} X_{ij}^2 = \dfrac{2}{n}\sum_{i=1}^{n}X_{ij}U_{ij} + \dfrac{1}{n}\sum_{i=1}^{n}U_{ij}^2 - \sigma_j^2.
$$
By condition \eqref{cond: MEsubGaussian} and properties of sub-Gaussian random variables, the variable
$
\sum_{i=1}^{n} X_{ij}U_{ij} 
$
is zero-mean sub-Gaussian with variance proxy $(\sum_{i=1}^{n}X_{ij}^2)\sigma_*^2$, which is of order $O(n)$ by condition \eqref{cond:boundonxsquare}. Therefore, by the Hoeffding inequality, there exists positive constants $d_1$ and $K_1$ such that
$$
\mathbb{P}\left(\left\vert \dfrac{1}{n} \sum_{i=1}^{n}X_{ij}U_{ij} \right \vert \geq d_1 \right)  = O\left(\exp\left(-K_1n\right)\right).
$$
Similarly, by condition \eqref{cond: MEsubGaussian}, the variables $U_{ij}^2$'s are sub-exponential with (finite) parameter ($\nu_0, \alpha_0)$, hence $\sum_{i=1}^{n}U_{ij}^2$ is also sub-exponential with parameter $(\nu_0 \sqrt{n}, \alpha_0)$. Because $\mathbb{E}(U_{ij}^2) = \sigma_j^2$, there exist constants $d_2 \in [0, \nu_0^2/\alpha_0]$ and $K_2$ such that
$$
\mathbb{P}\left(\left\vert \dfrac{1}{n} \sum_{i=1}^{n}U_{ij}^2 - \sigma_j^2 \right \vert \geq d_2 \right)  = O\left(\exp\left(-K_2n\right)\right).
$$
Combining these results, for some positive constants $d_3$ and $K_3$, we have
\begin{equation}
	P \left(\left\vert\dfrac{1}{n} \sum_{i=1}^{n}W_{ij}^2 - \sigma_j^2 - \dfrac{1}{n} \sum_{i=1}^{n} X_{ij}^2 \right\vert \geq d_3 \right) = O(\exp(-K_3n)).
	\label{eq: denominator}
\end{equation}
By condition \eqref{cond:boundonxsquare}, the term $n^{-1}\sum_{i=1}^{n} X_{ij}^2 = O(1)$, so inequality \eqref{eq: denominator} implies that $n^{-1}\sum_{i=1}^{n}W_{ij}^2 - \sigma_j^2$ is both bounded above and bounded away from zero with probability tending to one as $ n\to \infty$. Now for the numerator of $\eqref{eq:tildebetaj}$, we have
$$
\mathbb{E}\left(\dfrac{1}{n} \sum_{i=1}^{n}W_{ij}y_i\right) = \mathbb{E}\left(\dfrac{1}{n} \sum_{i=1}^{n}X_{ij}y_i\right) = \xi_{nj}, 
$$ and
$$
\dfrac{1}{n} \sum_{i=1}^{n}W_{ij}y_i - \xi_{nj} = \dfrac{1}{n}\sum_{i=1}^{n}X_{ij}\varepsilon_i + \dfrac{1}{n} \sum_{j=1}^{n} U_{ij}\varepsilon_i.
$$
By conditions \eqref{cond:epsilonsubGaussian} and \eqref{cond: MEsubGaussian}, using similar arguments, we obtain
$$
\mathbb{P}\left(\left\vert \dfrac{1}{n} \sum_{i=1}^{n}W_{ij}y_i - \xi_{nj} \right\vert \geq d_4 \right) =  O(\exp(-K_4n)),
$$
for some positive constant $d_4$ and $K_4$. Hence if $j \in \mathcal{S}$, along with the condition \eqref{cond:boundonxandy}, the above inequality implies that $\vert n^{-1} \sum_{i=1}^{n} W_{ij}y_i \vert$ is bounded away from zero with probability at least $1-O(\exp(-Cn))$. Using the union bound, there exist positive constants $d_7 = d_6/d_5$ with $ d_5 =  n^{-1}\sum_{i=1}^{n}X_{ij}^2 + d_3$ and $d_6 = \xi_{nj} - d_4$ such that for $j \in \mathcal{S}$,
\begin{align}
	\mathbb{P}(\vert\tilde{\beta_j} \vert \geq d_7) &  \geq \mathbb{P}\left(\left|\dfrac{1}{n} \sum_{i=1}^{n}W_{ij}y_i \right| \geq d_6 \text{ and }  \left|\dfrac{1}{n} \sum_{i=1}^{n}W_{ij}^2 - \sigma_j^2 \right| \leq d_5 \right) \nonumber \\ \nonumber & = 1 - \mathbb{P}\left(\left|\dfrac{1}{n} \sum_{i=1}^{n}W_{ij}y_i \right| \leq  d_6 \text{ or }  \left|\dfrac{1}{n} \sum_{i=1}^{n}W_{ij}^2 - \sigma_j^2 \right| \geq d_5  \right) 
	\nonumber 
	\\ & \geq  1 - \left\{ \mathbb{P}\left(\left|\dfrac{1}{n} \sum_{i=1}^{n}W_{ij}y_i \right| \leq  d_6 \right) + \mathbb{P} \left(\left|\dfrac{1}{n} \sum_{i=1}^{n}W_{ij}^2 - \sigma_j^2 \right| \geq d_5  \right)       \right\} \nonumber \\
	& = 1 - (O\exp(-K_4n)  + O(\exp(-K_3n))) = 1- O(\exp(-K_5n)),
	\label{eq:betakboundedbelow}
\end{align}
for some constant $C$. The final inequality states that for all $j \in \mathcal{S}$ we have $\vert \hat{\beta}_j \vert$ is bounded away from 0 with probability tending to one as $n \to \infty$.

Next, we will bound the norm $\lVert \tilde{\bm{\beta}} \rVert_2^2$ from above. Let $\bm{\Lambda}$  denote the $p\times p$ diagonal matrix with elements $n^{-1} \sum_{i=1}^{n} W_{ij}^2 - \sigma_j^2 ~j = 1,\ldots, p$, then we can write $\tilde{\bm{\beta}}$ as
$$
\tilde{\bm{\beta}} = \bm{\Lambda}^{-1} \left(\dfrac{1}{n} \bW^{\top} \mathbf{y} \right) = \bm{\Lambda}^{-1}\left(\dfrac{1}{n}\bX^\top \bX \bm{\beta}_0 +  \dfrac{1}{n}\bU^\top \bX \bm{\beta}_0 + \dfrac{1}{n}\bX^\top \bm{\varepsilon} + \dfrac{1}{n}\bU^\top \bm{\varepsilon} \right).
$$
Hence, by the submultiplicity of matrix norm and the triangle inequality, we have
\begin{equation}
	\lVert \tilde{\bm{\beta}} \rVert_2 \leq \Vert \bm\Lambda^{-1} \Vert_2 \left(\left\Vert\dfrac{1}{n}\bX^\top \bX \bm{\beta}_0 \right\Vert_2 +  \left\Vert \dfrac{1}{n}\bU^\top \bX \bm{\beta}_0 \right\Vert_2 +  \left\Vert \dfrac{1}{n}\bX^\top \bm{\varepsilon} \right\Vert_2 + \left\Vert \dfrac{1}{n}\bU^\top \bm{\varepsilon} \right\Vert_2\right)
	\label{eq:betatildenorm}
\end{equation}
where  $\left\Vert \bm\Lambda^{-1} \right\Vert_2 = \max_{j=1, \ldots, p}\left\{1/(\vert n^{-1} \sum_{i=1}^{n} W_{ij}^2 - \sigma_j^2 \vert )\right\} = \left\{\min_{j=1, \ldots, p} \vert n^{-1} \sum_{i=1}^{n} W_{ij}^2 - \sigma_j^2 \vert \right\}^{-1}$ is the magnitude of the largest eigenvalue of $\bm\Lambda^{-1}$ . From equation \eqref{eq: denominator} and condition \eqref{cond:boundonxsquare}, there exist sufficiently large constants $d_8$ and $K_8$ such that
\begin{align}
	\mathbb{P}\left(\left\Vert \bm\Lambda^{-1} \right\Vert_2 \leq d_8 \right) & = \mathbb{P} \left(\min_{j=1,\ldots,p} \left\{\left\vert \dfrac{1}{n} \sum_{i=1}^{n}W_{ij}^2 - \sigma_j^2\right\vert \right\} \geq \dfrac{1}{d_8}\right)
	\nonumber \\
	& =  \mathbb{P}\left(\bigcap\limits_{j= 1,\ldots, p}  \left\vert \dfrac{1}{n} \sum_{i=1}^{n}W_{ij}^2 - \sigma_j^2\right\vert \geq \dfrac{1}{d_8} \right) \nonumber\\ 
	& = 
	1 - \mathbb{P}\left(\bigcup\limits_{j= 1,\ldots, p}  \left\vert \dfrac{1}{n} \sum_{i=1}^{n}W_{ij}^2 - \sigma_j^2\right\vert \leq \dfrac{1}{d_8}  \right) \nonumber \\
	& \geq 1 - \sum_{j=1}^{p} \mathbb{P}\left(\left\vert \dfrac{1}{n} \sum_{i=1}^{n}W_{ij}^2 - \sigma_j^2\right\vert \leq \dfrac{1}{d_8}  \right) \nonumber \\ & = 1-O(p\exp(-K_8n)).
	\label{eq:betatildenorm_term0}
\end{align} 
Since $\log(p)/n \to C_1 $ by condition \eqref{cond:np}, as long as $K_8 > C_1$, the operator norm $\left\Vert \bm\Lambda^{-1} \right\Vert_2$ is bounded above by a constant with probability tending to one $n\to \infty$.  

Next, we will bound each term in the bracket of \eqref{eq:betatildenorm}. First, 
\begin{equation}
	\left\Vert\dfrac{1}{n}\bX^\top \bX \bm{\beta}_0 \right\Vert_2^2 = \bbeta_0^\top \left(\dfrac{1}{n}\bX^\top \bX \right)^2 \bbeta_0 \leq \lambda_{\max }^2\left(\dfrac{1}{n}\bX^\top \bX \right) \Vert\bbeta_0\Vert_2^2 = O(s n^{2\tau_2})
	\label{eq:betatildenorm_term1}
\end{equation}
by conditions \eqref{cond:boundonbetak} and \eqref{cond:boundonxsquare}. For the second term, 
\begin{align}
	\left\Vert\dfrac{1}{n}\bU^\top \bX \bm{\beta}_0 \right\Vert_2^2  & = \dfrac{1}{n^2}\bbeta_0^\top \bX^\top \bU \bU^\top \bX \bbeta_0  = \dfrac{1}{n^2} \bbeta_0^\top \bX^\top \bZ \bm{\Sigma}_u \bZ^\top \bX \bbeta_0 \nonumber \\ & 
	\leq \nonumber \dfrac{1}{n^2}\lambda_{\max }(\bm{\Sigma}_u) {\bbeta}_0^\top \bX^\top \bZ  \bZ^\top  \bX \bbeta_0 \nonumber \\
	& \leq \dfrac{1}{n^2} p\lambda_{\max }(\bm{\Sigma}_u) \lambda_{\max }(p^{-1}\bZ\bZ^\top)\bbeta_0^\top  \bX^\top \bX \bbeta_0 \nonumber \\
	& \leq  p \lambda_{\max }(\bm{\Sigma}_u) \lambda_{\max }(p^{-1}\bZ\bZ^\top) \lambda_{\max }\left(\dfrac{1}{n} \bX^\top \bX\right) \Vert \bbeta_0\Vert_2^2 \nonumber\\
	&  \stackrel{(i)}{\leq} pn^{\tau_1 +\tau_2-1} \lambda_{\max }(p^{-1}\bZ\bZ^\top) \Vert \bbeta_0\Vert_2^2 \nonumber,
\end{align}
where step $(i)$ follows conditions  \eqref{cond:MEeigenvalue} and \eqref{cond:boundoneigX}. Therefore, by conditions \eqref{cond:MEconcentration} and \eqref{cond:boundonbetak} there exist positive constants $d_9$ and $K_9$ such that 
$$
\mathbb{P}\left(\left\Vert\dfrac{1}{n}\bU^\top \bX \bm{\beta}_0 \right\Vert_2^2 \geq d_9psn^{\tau_1 + \tau_2-1})  \right)
\leq O(\exp(-K_9n)).
$$
For the third term, we have
$$
\left\Vert \dfrac{1}{n}\bX^\top \bm{\varepsilon} \right\Vert_2^2 = \dfrac{1}{n}\bm\varepsilon^\top \left(\dfrac{1}{n}\bX^\top \bX \right) \bm \varepsilon \leq \lambda_{\max } \left(\dfrac{1}{n}\bX^\top \bX \right) \dfrac{1}{n}\Vert \bm\varepsilon \Vert_2^2.
$$
By condition \eqref{cond:epsilonsubGaussian}, each term $\varepsilon_i^2$ is  sub-exponential with finite parameters; furthermore, since $\varepsilon_1, \ldots, \varepsilon_n$ are mutually independent, there exist positive constants $d_{10}$ and $K_{10}$  such that 
\begin{equation}
	\mathbb{P}\left(\dfrac{1}{n}\Vert \bm\varepsilon \Vert_2^2 \geq d_{10}\right) = \mathbb{P}\left(\dfrac{1}{n}\sum_{i=1}^{n} \varepsilon_i^2 \geq d_{10} \right)  = O(\exp(-K_{10}n)).
	\label{eq: normepsilon2}
\end{equation}
As a result, we obtain $$
\mathbb{P}\left(\left\Vert \dfrac{1}{n}\bX^\top \bm{\varepsilon} \right\Vert_2^2 \geq d_{11}n^{\tau_2}  \right) = O(\exp(-K_{11}n))
$$
for some positive constants $d_{11}$ and $K_{11}$. Finally, for the last term in the bracket of \eqref{eq:betatildenorm}, we have
\begin{align}
	\left\Vert \dfrac{1}{n}\bU^\top \bm{\varepsilon} \right\Vert_2^2 & = \dfrac{1}{n^2}\bm\varepsilon^\top \bU\bU^\top\bm{\varepsilon} = \dfrac{1}{n^2} \bm\varepsilon^\top \bZ \bm{\Sigma}_u\bZ^\top\bm{\varepsilon} \nonumber \\ & \leq  \dfrac{1}{n^2} \lambda_{\max }(\bm{\Sigma}_u) \bm\varepsilon^\top  \bZ\bZ^\top\bm{\varepsilon} \nonumber \\& \leq \dfrac{1}{n}p\lambda_{\max }(\bm{\Sigma}_u) \lambda_{\max }(p^{-1}\bZ\bZ^\top) \left(\dfrac{1}{n}\Vert\bm\varepsilon\Vert_2^2\right) \nonumber \\ & \leq c_3pn^{\tau_1-1}\lambda_{\max }(p^{-1}\bZ\bZ^\top) \left(\dfrac{1}{n}\Vert\bm\varepsilon\Vert_2^2\right).
\end{align}
Let $d_{12} = c_2c_3d_{10}$, then by condition \eqref{cond:MEconcentration} and \eqref{eq: normepsilon2}, we obtain
\begin{align}
	\mathbb{P}\left(\left\Vert \dfrac{1}{n}\bU^\top \bm{\varepsilon} \right\Vert_2^2 \geq d_{12} pn^{\tau_1-1} \right) & \leq \mathbb{P}\left\{ \lambda_{\max }(p^{-1}\bZ\bZ^\top) \left(\dfrac{1}{n}\Vert\bm\varepsilon\Vert_2^2\right) \geq c_2 d_{10} \right\}  \nonumber \\
	& = 1 - \mathbb{P}\left\{ \lambda_{\max }(p^{-1}\bZ\bZ^\top) \left(\dfrac{1}{n}\Vert\bm\varepsilon\Vert_2^2\right) \leq c_2 d_{10} \right\} \nonumber \\
	&\leq  1 - \mathbb{P}\left\{ \lambda_{\max }(p^{-1}\bZ\bZ^\top) \leq c_2 \text{ and } \dfrac{1}{n}\Vert\bm\varepsilon\Vert_2^2 \leq  d_{10} \right\} \nonumber \\
	& = \mathbb{P}\left\{ \lambda_{\max }(p^{-1}\bZ\bZ^\top) \geq c_2 \text{ or } \dfrac{1}{n}\Vert\bm\varepsilon\Vert_2^2 \geq  d_{10} \right\} \nonumber \\
	& \leq \mathbb{P}\left(\lambda_{\max }(p^{-1}\bZ\bZ^\top) \geq c_2 \right) + \mathbb{P}\left(\dfrac{1}{n}\Vert\bm\varepsilon\Vert_2^2 \geq  d_{10} \right) \nonumber \\ & = O(\exp(-K_{12}n))
	\label{eq:betatildenorm_term4}
\end{align}
for some positive constant $K_{12}$. Substituting \eqref{eq:betatildenorm_term0}-\eqref{eq:betatildenorm_term4} to \eqref{eq:betatildenorm}, for some sufficiently large constants $d_{13}$ and $K_{13} > C_1$ and under the condition that $p \geq n^{\tau_2 - \tau_1 +1}$, we have 

\begin{equation}
	\mathbb{P}\left(\Vert\tilde{\bbeta} \Vert_2^2 \geq d_{13}psn^{\tau_1+\tau_2 +  - 1}) \right) \leq   O(p\exp(-K_{13}n))
	\label{eq:betatildenorm_boundedabove}
\end{equation}
Finally, by Bonferonni's inequality, it follows from \eqref{eq:betakboundedbelow} and \eqref{eq:betatildenorm_boundedabove} that
\begin{equation}
	\mathbb{P} \left(\min_{i \in \mathcal{S}} \vert \tilde{\beta}_i \vert < d_{7} \text{ or } \lVert \tilde{\bm{\beta}} \rVert_2^2 > d_{13}psn^{\tau_1 + \tau_2 - 1}\right) \leq O\left(p \exp\left(-K_{14}n \right) \right),
	\label{eq:lemma_SISc}
\end{equation}
for some constants $K_{14} > C_1$. Hence, for sufficiently large constants $C>C_1$, with probability $1-O\left(p\exp\left(-Cn\right) \right)$, the magnitudes of $\tilde{\beta}_i, i \in \mathcal{S}$ are bounded away from zero and for some $d >0$, 
$$
\text{card}\left\{ 1 \leq k \leq p: \vert \tilde{\beta}_k \vert \geq \min_{i \in \mathcal{S}} \vert \tilde{\beta}_i \vert \right\} \leq d pn^{\tau_1 + \tau_2 + \log_n(s) - 1}. $$
Therefore, if $\delta \to 0$ and  satisfies $\delta n^{1-\tau_1-\tau_2 - \log_n(s)} \to \infty$ as $n \to \infty$, then \eqref{eq: main_SISc} holds with a constant $C > 0$ larger than $K_{14}$ in \eqref{eq:lemma_SISc}.

\paragraph{Step 2.} This step follows Step 2 in the proof of Theorem 1 of Fan \& Lv (2007). In this step, we will use $C$ to denote generic constants that are larger than $C_1$ in condition \eqref{cond:np}.  Fix an arbitrary $r \in (0,1)$ and choose a shrinking factor $\delta$ of the form $\delta = (\frac{n}{p})^{1/(k-r)}$ for some integer $k\geq 1$. We successively perform dimension reduction until the number of remaining
variables drops to below the sample size $n$. In other words  we obtain a sequence of nested sets $$
\widetilde{\mathcal{M}}_\delta^{k} \subset \widetilde{\mathcal{M}}_\delta^{k-1} \subset \ldots \subset \widetilde{\mathcal{M}}_\delta^1, 
$$
where each set $\widetilde{\mathcal{M}}_\delta = \widetilde{\mathcal{M}}_\delta^{j}$ has cardinality $\ceil{\delta^{j}p}$ and $d = \ceil{\delta^{k}p} = \ceil{\delta^{r}n} < n$ but $\ceil{\delta^{k-1}p} = \ceil{\delta^{r-1}n} > n$. Hence we see that $\widetilde{\mathcal{M}}_\delta = \hat{\mathcal{Q}}_\gamma$, with $\gamma = \delta^r < 1$.

Next, fix an arbitrary $\theta_1 \in (0, 1-\tau_1-\tau_2- \log_n(s))$ and pick some $r <1$ very close to 1 such that $\theta_0 = \theta_1/r < 1-\tau_1-\tau_2- \log_n(s)$, and choose a sequence of integers $k \geq 1$ in a way such that when $n \to \infty$,
\begin{equation}
	\delta n^{1-\tau_1-\tau_2-\log_n(s)} \to \infty \text{ and } \delta n^{\theta_0} \to 0, 
	\label{eq:40}
\end{equation}
where $\delta = (\frac{n}{p})^{1/(k-r)}$. Therefore, with the above dimension reduction process, we can raise both sides of \eqref{eq:40} to the $r$th power, and hence the set $\widetilde{\mathcal{M}}_\delta = \hat{\mathcal{Q}}_\gamma$ with $\gamma = \delta^r$ satisfies
$$
\gamma n^{r(1-\tau_1-\tau_2 -\log_n(s))} \to \infty \text{ and } \gamma n^{\theta_1} \to 0 .
$$
Since for any principal submatrix $\bm{\Sigma}_u^{0}$ of $\bm\Sigma_u$, we have $\lambda_{\max }(\bm{\Sigma}_u^{0}) < \lambda_{\max }(\bm{\Sigma}_u) \leq c_2n^{\tau_1}$, and that property \eqref{cond: concentrationproperty} in condition \eqref{cond:MEconcentration} holds for any $n \times \tilde{p}$ submatrix $\widetilde{\mathbf{Z}}$ of $\mathbf{Z}$ with $c_1n < \tilde{p} \leq p$, for some constant $C > 0$, in each step $1 \leq i \leq k$ in the above dimension reduction framework, we have
$$
\mathbb{P}\left(\widetilde{\mathcal{M}}_{\delta}^{i} \supseteq \mathcal{S}  \mid \widetilde{\mathcal{M}}_{\delta}^{i-1} \supseteq \mathcal{S} \right)=1-O\left(p\exp \left(-C n\right)\right),
$$
and hence by Bonferroni's inequality we have
\begin{equation}
	\mathbb{P}\left(\hat{\mathcal{Q}}_{\gamma} \supseteq \mathcal{S}\right)= 1-O\left(kp \exp \left(-C n\right)\right).
	\label{eq:42}
\end{equation}
It follows from \eqref{eq:40} that we require $\delta \to 0$ for all $r$ when both $n \to \infty$ and $p\ \to \infty$ (i.e $\log(n)/k - \log(p)/k \to -\infty)$. Since $p >n$ implies $k = O(\log(p)/\log(n))$, a suitable increase of the constant $C>0$ in \eqref{eq:42} gives
\begin{equation*}
	\mathbb{P}\left(\hat{\mathcal{Q}}_{\gamma}\supseteq \mathcal{S} \right)= 1-O\left(p \exp \left(-Cn\right)\right).
\end{equation*}
This  probability bound holds for any $\gamma \sim cn^{-\tilde\theta}$, with $\tilde\theta < 1-\tau_1-\tau_2 - \log_n(s)$ and $c>0$, completing the proof.

\section{Preprocessing for the microarray data}
In this subsection, we provide more detail on the steps that we used to preprocess the microarray data in Section 4 of the main paper. 

As noted in \citet{hein2005bgx}, on Affymetrix ``GeneChips'' oligonucleotide arrays, each gene is represented by a probe set, consisting of a number of probe pairs. A probe pair further contains a perfect match (PM) probe and a mismatch probe (MM). On the one hand, the intensity observed for the PM measurement for probe $r$ at gene $j$ was assumed due partly to the binding of the cRNA that perfectly matches the sequence on the array, denoted as $S_{jr}$ (true signal), and partly to hybridization of the cRNA that did not perfectly match the sequence, denoted as $H_{jr}$ (nonspecific hybridization). On the other hand, the intensity observed for the MM measurement for probe $r$ at gene $j$ was due partly to binding of a fraction $\phi \in (0,1)$ of the true signal and partly due to nonspecific hybridization. Furthermore, it was assumed that
\begin{align*}
	\text{PM}_{jr} & \sim N(S_{jr} + H_{jr}, \tau^2), \\
	\text{MM}_{jr} & \sim N(\phi S_{jr} + H_{jr}, \tau^2),
\end{align*}
where $N(\mu, \sigma^2)$ denotes the normal distribution with mean $\mu$ and variance $\sigma^2$. Next, \citet{hein2005bgx} modeled the true signal $S_{gj}$ and nonspecific hybridization $H_{gj}$ on the log scale. Since they allowed them to be zero, they assumed that
\begin{align*}
	\log(S_{jr} + 1 ) \sim \text{TN}(X_j, \sigma_j^2), \\
	\log(H_{jr} + 1 ) \sim \text{TN}(\lambda, \eta^2),
\end{align*}
where $\text{TN}(\mu, \sigma^2)$ denotes the normal distribution with mean $\mu$ and variance $\sigma^2$ left truncated at 0. Furthermore, to account for outlying probes and stabilize the gene-specific variance parameters, all the variances $\sigma_g^2$ were assumed to be exchangeable,
$$
\log(\sigma_j^2) \stackrel{iid}{\sim} N(a, b^2).
$$
Finally, the Bayesian model was fully specified by assuming non-informative priors on $X_j$, $\phi$, $\lambda$, $\tau^{-2}$ and $\eta^{-2}$, while $a$ and $b$ were fixed at values obtained by an empirical procedure. The primary parameter of interest was $X_j$, the mean gene expression level for gene $j$ on the log scale. 

We fitted the above Bayesian model for each patient $i=1,\ldots, n$ separately by the BGX package of \citet{hein2005bgx} in R and obtained the posterior distribution for $X_{ij}$. Let
$\widehat{W}_{ij}$ and $\text{var}(\widehat{X}_{ij})$ denote the posterior mean and variance of $X_{ij}$ respectively; so we considered $\widehat{W}_{ij}$  as a surrogate for the true $X_{ij}$ contaminated by some measurement error $U_{ij}$ with variance $\text{Var}(\widehat{X}_{ij})$. Next, for each gene $j$, we standardized the posterior mean $\tilde{W}_{ij}$ to obtain $W_{ij} = {(\widehat{W}_{ij}-\overline{W}_j)}/{s_j}, \ i=1,\ldots,n$, where $\overline{W}_j = n^{-1} \sum_{i=1}^{n}\widehat{W}_{ij}$ and $s_j^2 = n^{-1} \sum_{i=1}^{n} (\widehat{W}_{ij}-\overline{W}_j)^2$.
To simplify the analysis, we assumed that measurement error was independent of the patient's true gene expression levels and that the associated variance was constant across patients for a given gene; hence, the matrix $\bm{\Sigma}_u$ was set to be diagonal. We averaged the posterior variance, $\widehat{\sigma}^2_{u,j}= n^{-1} \sum_{i=1}^{n}\text{Var}(\widehat{X}_{ij})$, and  the diagonal elements of the measurement error covariance matrix  $\bm{\Sigma}_u$ were estimated as $(\widehat{\bm{\Sigma}}_u)_{j,j} = {\widehat{\sigma}^2_{uj}}/s^2_j, j=1,\ldots,p$. As noted by \citet{sorensen2015measurement}, if the measurement error variance is too large compared to the between-sample variance of the between-patient variability, then little can be done. Hence, only the $p=993$ genes with $\widehat{\sigma}^2_{u,j} < 0.5 s_j^2$, i.e. estimated noise-to-signal ratio less than $1$, were retained for further analysis.

\section{Efficiency of the screening procedures}
In this section, we present additional simulation results to investigate 
the efficiency of the proposed screening procedures. We generated data in the same way as in the subsection 3.1 of the main paper, with the exception that we considered one more scenario for $\bm\Sigma_u$, the covariance matrix of the measurement error and set $\rho_x = 0.5$. In this new scenario, $\bm\Sigma_u$ was set to have a homogeneous structure, with all off-diagonal elements equal to $0.2$ and diagonal elements equal to $0.4$; hence the measurement errors on all covariates were correlated with one another.

\begin{figure}
	\centering
	\begin{subfigure}{.8\textwidth}
		\centering
		\caption{$\bm{\Sigma}_x$ has an homogeneous structure.}
		\includegraphics[width = \textwidth]{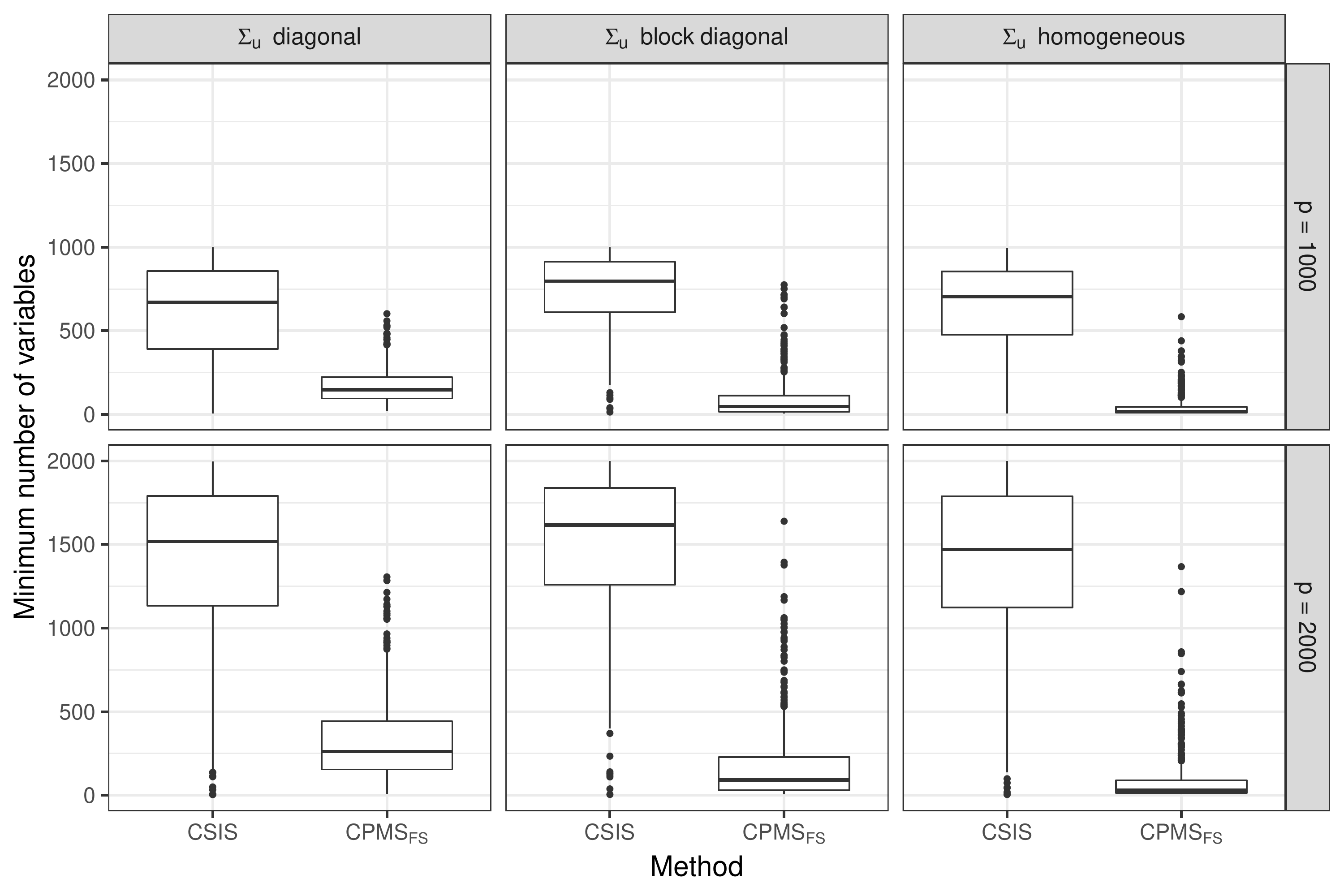}
	\end{subfigure}\vspace{0.05\textwidth} 
	\begin{subfigure}{.8\textwidth}
		\caption{$\bm{\Sigma}_x$ has an AR(1) structure.}
		\centering
		\includegraphics[width = \textwidth]{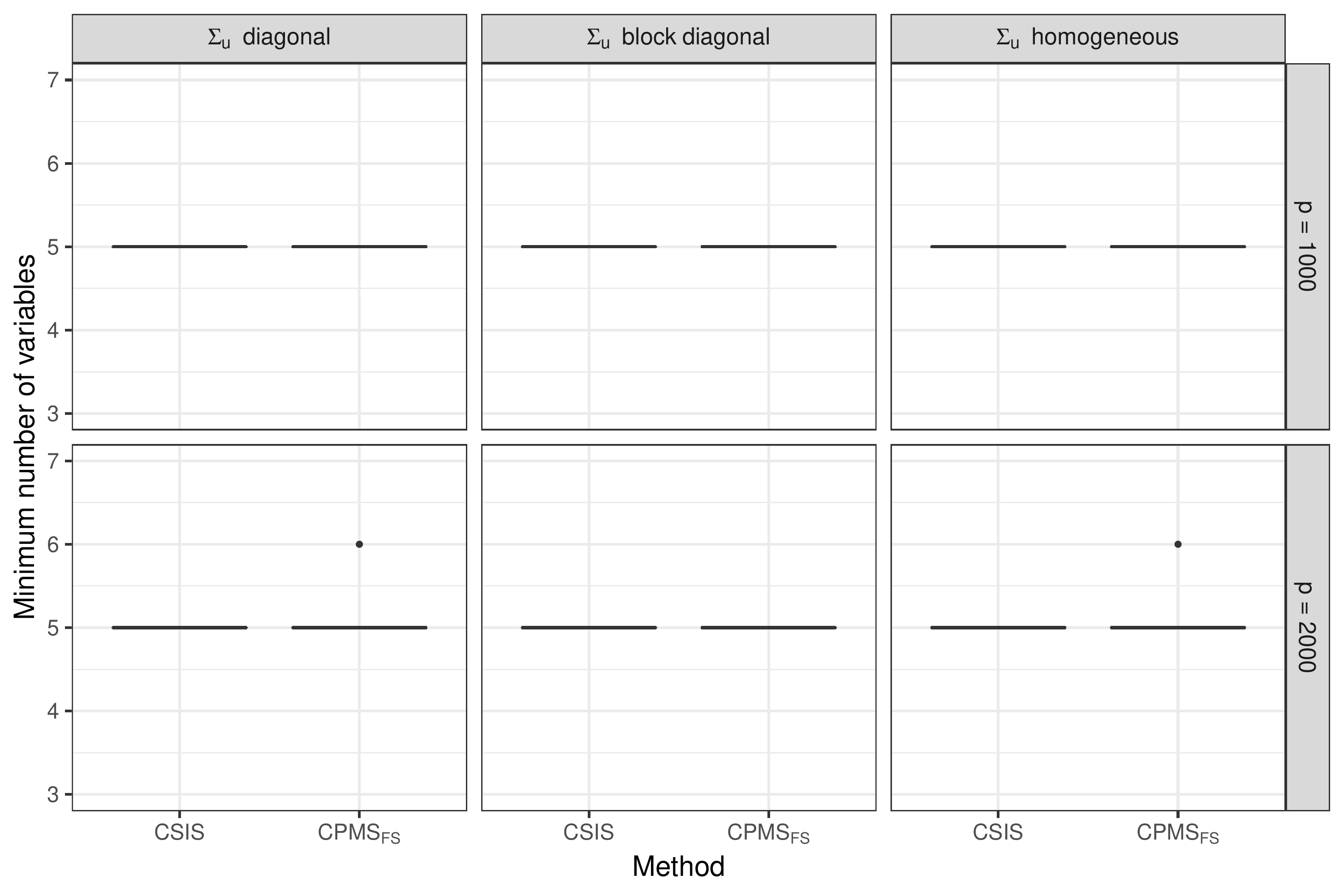}
	\end{subfigure}
	\caption{Minimum number of variables that has to be kept for the SISc and PMSc$_{\text{FS}}$ to retain all the important variables in the simulation study.}
	\label{fig:min_CS}
\end{figure}

We report the mean false positive rate (FPR) and mean false negative rate (FNR) across $500$ samples for PMSc$_{\text{CV}}$ (Table \ref{table:pmb_cv}), and the empirical cumulative distribution of the \textit{minimum} number of variables that had to be included for PMSc forward stepwise and SISc in order to retain all the important variables (Figure \ref{fig:min_CS}). Note that the latter is always between $s$ and $p$. On the one hand, a low minimum number of variables to be included indicates a high screening efficiency; on the other hand, if the minimum number of variables to be included is close to $p$, then dimension reduction is not achievable via screening.

\begin{table}[ht]
	\centering
	\caption{Average false positive rates (FPR, in percentage) and false negative rates (FNR, in percentage) of PMSc$_\text{CV}$ with tuning parameters selected via five fold cross-validation. Standard errors are included in parentheses.}
	\resizebox{\textwidth}{!}{\begin{tabular}{rrrrrrrr}
			\toprule[1.5pt]
			$\bm{\Sigma}_x$ & $p$ & \multicolumn{2}{c}{$\bm{\Sigma}_u$ diagonal} &
			\multicolumn{2}{c}{$\bm{\Sigma}_u$ block diagonal} 
			& \multicolumn{2}{c}{$\bm{\Sigma}_u$ homogeneous}\\  
			&&	FPR  & FNR & FPR & FNR & FPR & FNR  \\ 
			\cmidrule{3-8}
			
			AR(1) & 1000 & 0.1 (0.1) & 0.0 (0.0) & 0.0 (0.1) & 0.0 (0.0) & 0.1 (0.1) & 0.0 (0.0) \\ 
			& 2000 & 0.0 (0.0) & 0.0 (0.0) & 0.0 (0.0) & 0.0 (0.0) & 0.0 (0.0) & 0.0 (0.0) \\ 
			\multirow{2}{1.5cm}{Homogen} & 1000 & 100.0 (0.0) &   0.0 (0.0) & 100.0 (0.0) &  0.0 (0.0) & 100.0 (0.0) &   0.0 (0.0) \\ 
			& 2000 & 100.0 (0.0) &   0.0 (0.0) & 100.0 (0.0) &  0.0 (0.0) & 100.0 (0.0) &   0.0 (0.0)  \\ 
			\bottomrule[1.5pt]
	\end{tabular}}
	\label{table:pmb_cv}
\end{table}

Table \ref{table:pmb_cv} and Figure \ref{fig:min_CS} indicate that when $\bm{\Sigma}_x$ had an AR(1) structure, all the screening methods had high efficiency in variable selection. Specifically, regardless of the structure of measurement error variance $\bm{\Sigma}_u$, the PMSc$_{\text{CV}}$ method had both false positive and false negative rates very close to zero. For the SISc and PMSc$_{\text{FS}}$, the number of variables each had to keep was also close to $5$, the true number of non-zero coefficients in $\bbeta_0$. These results confirm both the variable selection consistency of PMSc and the ability of SISc to reduce the number of dimensions dramatically when all the true covariates are not highly collinear. 

By contrast, when $\bm{\Sigma}_x$ had a homogeneous structure, the efficiency of screening methods decreased considerably. The PMSc$_{\text{CV}}$ method was essentially unable to reduce the number of variables in that case, as evidenced by the false positive rates close to 100\%. Similarly, to ensure all the important variables were retained, the SISc method had to keep 75-80\% of the variables, which was usually more than the sample size $n$. Among the three screening methods, the PMSc$_{\text{FS}}$ was the most effective; in all the considered settings, it could effectively reduce the dimension to below the sample size $n = 500$. Looking at the distribution of the minimum number of variables that must be kept for PMSc$_{\text{FS}}$ more closely, it is interesting that the efficiency of PMSc$_{\text{FS}}$ seemed to increase when measurement errors were more correlated (i.e $\bm\Sigma_u$ is homogeneous).
\newpage
\section{Detailed simulation results for one-stage and two-stage estimators} 
\begin{table}[ht]
	\centering
	\caption{Performance of the one-stage and two-stage estimators in the simulation study based on false positive rate (FPR, in percentage), false negative rate (FNR, in percentage), $\ell_2$ estimation error, and computation time (in seconds) when $\bm{\Sigma}_x$ has an AR(1) structure with autocorrelation $\rho_x = 0.5$. Standard error are included in parentheses. The CoCo estimator was only computed when the number of variables was no more than $1000$ (either in the 1st or 2nd step).}
	\medskip
	\resizebox{\linewidth}{!}{\begin{tabular}{lll rr rrr rr}
			\toprule[1.5pt]
			$\bm\Sigma_u$  & $p$ & Estimator &  \multicolumn{2}{c}{1st step} &
			\multicolumn{3}{c}{2nd steps} & \multicolumn{2}{c}{Time}\\
			& & & FPR& FNR & FPR & FNR  & $\ell_2$& 1st step & 2nd step \\ 
			\cmidrule(lr){4-5} \cmidrule(lr){6-8} \cmidrule(lr){9-10}
			Diagonal & 1000 & One-stage Corrected & - &  - &  2.2 (1.1) &  0.0 (0.0) & 0.35 (0.10) &   0.0 (0.0) &  49.7 (19.5) \\ 
			&  & PMSc$_{\text{CV}}$-Corrected &  0.1 (0.1) &  0.0 (0.0) &  0.1 (0.1) &  0.0 (0.0) & 0.26 (0.11) &  23.4 (0.3) &   0.4 (0.1) \\ 
			&  & PMSc$_{\text{FS}}$-Corrected &   7.5 (0.0) &  0.0 (0.0) &  0.0 (0.0) &  0.0 (0.0) & 0.28 (0.10) &  5.4 (1.1) &  1.6 (0.5) \\ 
			&  & SISc-Corrected &  7.5 (0.0) &  0.0 (0.0) &  1.2 (0.6) &  0.0 (0.0) & 0.33 (0.10) &   1.5 (0.0) &   2.0 (0.6) \\ 
			&  & One-stage CoCo &  - &  - &  7.6 (2.0) &  0.0 (0.0) & 0.47 (0.11) &   0.0 (0.0) & 516.8 (32.1) \\ 
			&  & PMSc$_{\text{CV}}$-CoCo &  0.1 (0.1) &  0.0 (0.0) &  0.1 (0.0) &  0.0 (0.0) & 0.26 (0.11) &  23.4 (0.3) &   0.0 (0.0) \\ 
			&  & PMSc$_{\text{FS}}$-CoCo &  7.5 (0.0) &  0.0 (0.0) &  0.0 (0.0) &  0.0 (0.0) & 0.27 (0.09) &  5.4 (1.1) &  0.7 (0.2)\\ 
			&  & SISc-CoCo &  7.5 (0.0) &  0.0 (0.0) &  0.1 (0.1) &  0.0 (0.0) & 0.31 (0.09) &   1.5 (0.0) &   0.8 (0.2) \\[3pt]
			
			& 2000 & One-stage Corrected &  - &  - &  1.3 (0.6) &  0.0 (0.0) & 0.36 (0.10) &   0.0 (0.0) & 136.5 (55.7) \\ 
			&  & PMSc$_{\text{CV}}$-Corrected &  0.0 (0.0) &  0.0 (0.0) &  0.0 (0.0) &  0.0 (0.0) & 0.26 (0.10) &  38.0 (2.0) &   0.3 (0.1) \\ 
			&  & PMSc$_{\text{FS}}$-Corrected &  3.8 (0.0) &  0.0 (0.0) &  0.6 (0.3) &  0.0 (0.0) & 0.29 (0.10) &  17.2 ( 3.1) &   2.4 ( 0.7)  \\ 
			&  & SISc-Corrected &  3.8 (0.0) &  0.0 (0.0) &  0.6 (0.3) &  0.0 (0.0) & 0.33 (0.10) &   2.8 (0.1) &   1.8 (0.5) \\ 
			&  & PMSc$_{\text{CV}}$-CoCo &  0.0 (0.0) &  0.0 (0.0) &  0.0 (0.0) &  0.0 (0.0) & 0.26 (0.11) &  38.0 (2.0) &   0.0 (0.0) \\ 
			&  & PMSc$_{\text{FS}}$-CoCo &  3.8 (0.0) &  0.0 (0.0) &  0.1 (0.0) &  0.0 (0.0) & 0.27 (0.09) &  17.2 ( 3.1) &   0.8 ( 0.2) \\ 
			&  & SISc-CoCo &  3.8 (0.0) &  0.0 (0.0) &  0.1 (0.0) &  0.0 (0.0) & 0.31 (0.09) &   2.8 (0.1) &   0.7 (0.2) \\ \\
			
			\multirow{2}{1.2cm}{Block diagonal} & 1000 & One-stage Corrected &  - &  - &  1.8 (0.9) &  0.0 (0.0) & 0.62 (0.17) &   0.0 (0.0) &  27.8 (14.9) \\ 
			&  & PMSc$_{\text{CV}}$-Corrected &  0.0 (0.0) &  0.0 (0.0) &  0.0 (0.0) &  0.0 (0.0) & 0.43 (0.18) &  23.4 (0.4) &   0.4 (0.1) \\ 
			&  & PMSc$_{\text{FS}}$-Corrected &  7.5 (0.0) &  0.0 (0.0) &  0.0 (0.0) &  0.0 (0.4) & 0.48 (0.16) &  5.3 (1.0) &  1.9 (0.5)  \\ 
			&  & SISc-Corrected &  7.5 (0.0) &  0.0 (0.0) &  1.0 (0.5) &  0.0 (0.0) & 0.57 (0.18) &   1.5 (0.0) &   2.6 (0.8) \\ 
			&  & One-stage CoCo &  - &  - &  7.4 (1.9) &  0.0 (0.0) & 0.74 (0.13) &   0.0 (0.0) & 417.4 (22.2) \\ 
			&  & PMSc$_{\text{CV}}$-CoCo &  0.0 (0.0) &  0.0 (0.0) &  0.0 (0.0) &  0.0 (0.0) & 0.43 (0.18) &  23.4 (0.4) &   0.0 (0.0) \\ 
			&  & PMSc$_{\text{FS}}$-CoCo &  7.5 (0.0) &  0.0 (0.0) &  0.0 (0.0) &  0.0 (0.4) & 0.46 (0.13) &  4.9 (1.6) &  1.3 (0.2) \\ 
			&  & SISc-CoCo &  7.5 (0.0) &  0.0 (0.0) &  0.1 (0.0) &  0.0 (0.0) & 0.54 (0.15) &   1.5 (0.0) &   1.7 (0.2) \\[3pt] 
			
			& 2000 & One-stage Corrected &  - &  - &  1.0 (0.4) &  0.0 (0.9) & 0.64 (0.17) &  0.0 (0.0) & 68.0 (36.1) \\ 
			&  & PMSc$_{\text{CV}}$-Corrected &  0.0 (0.0) &  0.0 (0.0) &  0.0 (0.0) &  0.0 (0.0) & 0.44 (0.17) & 37.9 (1.6) &  0.3 (0.1) \\ 
			&  & PMSc$_{\text{FS}}$-Corrected &  3.8 (0.0) &  0.0 (0.0) &  0.5 (0.2) &  0.1 (1.3) & 0.46 (0.14) & 16.7 (3.6) &  2.8 (0.8) \\ 
			&  & SISc-Corrected &  3.8 (0.0) &  0.0 (0.0) &  0.5 (0.2) &  0.0 (0.0) & 0.57 (0.17) &  2.8 (0.1) &  2.3 (0.6) \\ 
			&  & PMSc$_{\text{CV}}$-CoCo &  0.0 (0.0) &  0.0 (0.0) &  0.0 (0.0) &  0.0 (0.0) & 0.45 (0.18) & 37.9 (1.6) &  0.0 (0.0) \\ 
			&  & PMSc$_{\text{FS}}$-CoCo &   3.8 (0.0) &  0.0 (0.0) &  0.1 (0.0) &  0.0 (0.9) & 0.44 (0.13) & 16.7 (3.6) &  1.7 (0.2)  \\ 
			&  & SISc-CoCo &  3.8 (0.0) &  0.0 (0.0) &  0.1 (0.0) &  0.0 (0.0) & 0.54 (0.15) &  2.8 (0.1) &  1.4 (0.2) \\ 
			\bottomrule[1.5pt]
	\end{tabular}}
	\label{tab:AR1Sigmax_0.5}
\end{table}

\begin{table}[ht]
	\centering
	\caption{Performance of the one-stage and two-stage estimators in the simulation study based on false positive rate (FPR, in percentage), false negative rate (FNR, in percentage), $\ell_2$ estimation error, and computation time (in seconds) when $\bm{\Sigma}_x$ has an homogeneous structure with $\rho_x = 0.5$. Standard error are included in parentheses. The CoCo estimator was only computed when the number of variables was no more than $1000$ (either in the 1st or 2nd step).}
	\resizebox{\linewidth}{!}{\begin{tabular}{lll ll rrr rr}
			\toprule[1.5pt]
			$\bm\Sigma_u$  & $p$ & Estimator &  \multicolumn{2}{c}{1st step} &
			\multicolumn{3}{c}{2nd step} & \multicolumn{2}{c}{Time}\\
			
			& & & FPR & FNR & FPR  & FNR & $\ell_2$ & 1st step & 2nd step \\ 
			\cmidrule(lr){4-5} \cmidrule(lr){6-8} \cmidrule(lr){9-10}
			Diagonal & 1000 & One-stage Corrected &   - &   - &  20.0 (29.0) &   0.1 (1.5) & 0.96 (0.77) &   0.0 (0.0) &  18.9 (3.3) \\ 
			&  & PMSc$_{\text{CV}}$-Corrected & 100.0 (0.0) &   0.0 (0.0) &  19.7 (28.8) &   0.2 (2.0) & 0.95 (0.77) &  12.7 (0.7) &  18.7 (3.3) \\ 
			&  & PMSc$_{\text{FS}}$-Corrected &   7.6 (0.0) &   3.2 (7.8) &   1.2 (1.6) &   3.2 (7.8) & 0.56 (0.41) &   5.4 (0.7) &   1.4 (0.2) \\ 
			&  & SISc-Corrected &   7.6 (0.1) &   6.7 (10.5) &   1.0 (0.9) &   6.7 (10.5) & 0.54 (0.30) &   1.3 (0.1) &   1.0 (0.2) \\ 
			&  & One-stage CoCo &   - &   - &  11.5 (2.2) &   0.0 (0.0) & 0.77 (0.13) &   0.0 (0.0) & 481.5 (27.5) \\ 
			&  & PMSc$_{\text{CV}}$-CoCo & 100.0 (0.0) &   0.0 (0.0) &  11.5 (2.2) &   0.0 (0.0) & 0.77 (0.13) &  12.7 (0.7) & 476.7 (26.8) \\ 
			&  & PMSc$_{\text{FS}}$-CoCo &   7.6 (0.0) &   3.2 (7.8) &   1.0 (0.4) &   3.2 (7.8) & 0.45 (0.18) &   5.4 (0.7) &   1.8 (0.3) \\ 
			&  & SISc-CoCo &   7.6 (0.1) &   6.7 (10.5) &   1.0 (0.4) &   6.7 (10.5) & 0.52 (0.23) &   1.3 (0.1) &   1.6 (0.2) \\[3pt] 
			
			& 2000 & One-stage Corrected &   - &   - &  29.1 (21.0) &   1.1 (5.1) & 1.57 (0.84) &  0.0 (0.0) & 49.9 (7.0) \\ 
			&  & PMSc$_{\text{CV}}$-Corrected & 100.0 (0.0) &   0.0 (0.0) &  28.3 (21.3) &   1.0 (5.1) & 1.53 (0.84) & 25.2 (1.4) & 49.3 (6.9) \\ 
			&  & PMSc$_{\text{FS}}$-Corrected &   3.8 (0.0) &   4.0 (8.5) &   0.7 (0.9) &   4.0 (8.5) & 0.62 (0.51) &  8.5 (1.2) &  1.1 (0.2)  \\ 
			&  & SISc-Corrected &   3.8 (0.0) &  11.0 (13.1) &   0.5 (0.4) &  11.0 (13.1) & 0.61 (0.31) &  2.7 (0.1) &  0.9 (0.2) \\ 
			&  & PMSc$_{\text{CV}}$-CoCo & 100.0 (0.0) &   0.0 (0.0) &   - &   - &  - & 25.2 (1.4) &  - \\ 
			&  & PMSc$_{\text{FS}}$-CoCo &   3.8 (0.0) &   4.0 (8.5) &   0.5 (0.2) &   4.0 (8.5) & 0.45 (0.18) &  8.5 (1.2) &  1.7 (0.2) \\ 
			&  & SISc-CoCo &   3.8 (0.0) &  11.0 (13.1) &   0.5 (0.3) &  11.0 (13.1) & 0.60 (0.29) &  2.7 (0.1) &  1.9 (0.2) \\ \\
			\multirow{2}{1.2cm}{Block diagonal} & 1000 & One-stage Corrected &   - &   - &  23.5 (29.0) &   1.1 (5.2) & 1.25 (0.73) &   0.0 (0.0) &  22.7 (4.5) \\ 
			&  & PMSc$_{\text{CV}}$-Corrected & 100.0 (0.0) &   0.0 (0.0) &  25.6 (29.4) &   0.8 (4.2) & 1.30 (0.74) &  14.5 (0.7) &  22.5 (4.6) \\ 
			&  & PMSc$_{\text{FS}}$-Corrected &   7.6 (0.0) &   3.4 (8.1) &   3.0 (3.2) &   3.6 (8.2) & 1.15 (0.69) &   5.3 (0.7) &   1.4 (0.3) \\ 
			&  & SISc-Corrected &   7.6 (0.1) &  14.2 (14.8) &   1.5 (2.1) &  14.2 (14.9) & 0.97 (0.48) &   1.3 (0.1) &   1.1 (0.3) \\ 
			&  & One-stage CoCo &   - &   - &  14.3 (2.3) &   0.4 (3.2) & 1.23 (0.17) &   0.0 (0.0) & 417.2 (28.2) \\ 
			&  & PMSc$_{\text{CV}}$-CoCo & 100.0 (0.0) &   0.0 (0.0) &  14.3 (2.2) &   0.5 (3.3) & 1.23 (0.17) &  14.5 (0.7) & 413.4 (28.5) \\ 
			&  & PMSc$_{\text{FS}}$-CoCo &   7.6 (0.0) &   3.4 (8.1) &   0.9 (0.4) &   3.6 (8.2) & 0.68 (0.18) &   5.3 (0.7) &   2.9 (0.3) \\ 
			&  & SISc-CoCo &   7.6 (0.1) &  14.2 (14.8) &   0.9 (0.4) &  14.2 (14.8) & 0.84 (0.31) &   1.3 (0.1) &   2.3 (0.2) \\[3pt]
			
			& 2000 & One-stage Corrected &   - &   - &  32.4 (17.1) &   3.0 (10.1) & 1.86 (0.65) &  0.0 (0.0) & 52.5 (9.7) \\ 
			&  & PMSc$_{\text{CV}}$-Corrected & 100.0 (0.0) &   0.0 (0.0) &  31.8 (17.4) &   3.3 (10.1) & 1.84 (0.65) & 27.7 (1.5) & 52.1 (9.6) \\ 
			&  & PMSc$_{\text{FS}}$-Corrected &   3.8 (0.0) &   6.6 (10.8) &   1.6 (1.6) &   6.6 (10.9) & 1.21 (0.69) &  8.3 (1.1) &  1.1 (0.2)  \\ 
			&  & SISc-Corrected &   3.8 (0.0) &  21.4 (16.6) &   0.8 (1.0) &  21.4 (16.6) & 1.09 (0.48) &  2.8 (0.1) &  1.0 (0.3) \\ 
			&  & PMSc$_{\text{CV}}$-CoCo & 100.0 (0.0) &   0.0 (0.0) &   - &   - &  - & 27.7 (1.5) &  - \\ 
			&  & PMSc$_{\text{FS}}$-CoCo &   3.8 (0.0) &   6.6 (10.8) &   0.5 (0.2) &   6.7 (10.9) & 0.73 (0.22) &  8.3 (1.1) &  2.5 (0.2)\\ 
			&  & SISc-CoCo &   3.8 (0.0) &  21.4 (16.6) &   0.5 (0.3) &  21.4 (16.6) & 0.97 (0.36) &  2.8 (0.1) &  2.5 (0.2) \\ 
			\bottomrule[1.5pt]
	\end{tabular}}
	\label{tab:HomogenSigmax_0.5}
\end{table}

\begin{table}[ht]
	\centering
	\caption{Performance of the one-stage and two-stage estimators in the simulation study based on false positive rate (FPR, in percentage), false negative rate (FNR, in percentage), $\ell_2$ estimation error, and computation time (in seconds) when $\bm{\Sigma}_x$ has an AR(1) structure with autocorrelation $\rho_x = 0.3$. Standard error are included in parentheses. The CoCo estimator was only computed when the number of variables was no more than $1000$ (either in the 1st or 2nd step).}
	\medskip
	\resizebox{\linewidth}{!}{\begin{tabular}{lll rr rrr rr}
			\toprule[1.5pt]
			$\bm\Sigma_u$  & $p$ & Estimator &  \multicolumn{2}{c}{1st step} &
			\multicolumn{3}{c}{2nd step} & \multicolumn{2}{c}{Time}\\
			
			& & & FPR & FNR & FPR  & FNR  & $\ell_2$ & 1st step & 2nd step \\ 
			\cmidrule(lr){4-5} \cmidrule(lr){6-8} \cmidrule(lr){9-10}	
			Diagonal & 1000 & One-stage Corrected &  - &  - &  2.5 (1.3) &  0.0 (0.0) & 0.35 (0.07) &   0.0 (0.0) &  31.5 (10.9) \\ 
			&  & PMSc$_{\text{CV}}$-Corrected &  0.0 (0.0) &  0.0 (0.0) &  0.0 (0.0) &  0.0 (0.0) & 0.19 (0.07) &  23.4 (0.3) &   0.3 (0.0) \\ 
			&  & PMSc$_{\text{FS}}$-Corrected &  7.5 (0.0) &  0.0 (0.0) &  0.0 (0.0) &  0.0 (0.0) & 0.25 (0.07) &  5.3 (1.0) &  1.3 (0.3)  \\ 
			&  & SISc-Corrected &  7.5 (0.0) &  0.0 (0.0) &  1.2 (0.6) &  0.0 (0.0) & 0.31 (0.07) &   1.5 (0.0) &   1.8 (0.3) \\ 
			&  & One-stage CoCo &  - &  - &  8.6 (1.9) &  0.0 (0.0) & 0.49 (0.08) &   0.0 (0.0) & 519.1 (33.7) \\ 
			&  & PMSc$_{\text{CV}}$-CoCo &  0.0 (0.0) &  0.0 (0.0) &  0.0 (0.0) &  0.0 (0.0) & 0.19 (0.07) &  23.4 (0.3) &   0.0 (0.0) \\ 
			&  & PMSc$_{\text{FS}}$-CoCo & 7.5 (0.0) &  0.0 (0.0) &  0.0 (0.0) &  0.0 (0.0) & 0.26 (0.07) &  5.3 (1.0) &  0.6 (0.1)  \\ 
			&  & SISc-CoCo &  7.5 (0.0) &  0.0 (0.0) &  0.2 (0.1) &  0.0 (0.0) & 0.32 (0.07) &   1.5 (0.0) &   0.6 (0.1) \\[3pt]
			
			& 2000 & One-stage Corrected &  - &  - &  1.5 (0.7) &  0.0 (0.0) & 0.36 (0.08) &  0.0 (0.0) & 80.6 (30.8) \\ 
			&  & PMSc$_{\text{CV}}$-Corrected &  0.0 (0.0) &  0.0 (0.0) &  0.0 (0.0) &  0.0 (0.0) & 0.19 (0.07) & 38.2 (1.6) &  0.2 (0.1) \\ 
			&  & PMSc$_{\text{FS}}$-Corrected &  3.8 (0.0) &  0.0 (0.0) &  0.6 (0.3) &  0.0 (0.0) & 0.26 (0.07) &  16.9 (2.9) &   2.0 (0.4) \\ 
			&  & SISc-Corrected &  3.8 (0.0) &  0.0 (0.0) &  0.6 (0.3) &  0.0 (0.0) & 0.31 (0.08) &  2.8 (0.1) &  1.5 (0.3) \\ 
			&  & PMSc$_{\text{CV}}$-CoCo &  0.0 (0.0) &  0.0 (0.0) &  0.0 (0.0) &  0.0 (0.0) & 0.19 (0.07) & 38.2 (1.6) &  0.0 (0.0) \\ 
			&  & PMSc$_{\text{FS}}$-CoCo &  3.8 (0.0) &  0.0 (0.0) &  0.1 (0.1) &  0.0 (0.0) & 0.27 (0.07) &  16.9 (2.9) &   0.6 (0.1)  \\ 
			&  & SISc-CoCo &  3.8 (0.0) &  0.0 (0.0) &  0.1 (0.1) &  0.0 (0.0) & 0.31 (0.07) &  2.8 (0.1) &  0.6 (0.1) \\ \\
			
			\multirow{2}{1.2cm}{Block diagonal} & 1000 & One-stage Corrected &  - &  - &  2.0 (0.8) &  0.0 (0.0) & 0.59 (0.12) &   0.0 (0.0) &  18.9 (7.4) \\ 
			&  & PMSc$_{\text{CV}}$-Corrected &  0.0 (0.0) &  0.0 (0.0) &  0.0 (0.0) &  0.0 (0.0) & 0.36 (0.13) &  23.4 (0.4) &   0.3 (0.1) \\ 
			&  & PMSc$_{\text{FS}}$-Corrected &  7.5 (0.0) &  0.0 (0.0) &  0.0 (0.0) &  0.0 (0.2) & 0.45 (0.13) &  5.4 (1.1) &  1.6 (0.3)  \\ 
			&  & SISc-Corrected &  7.5 (0.0) &  0.0 (0.0) &  1.0 (0.5) &  0.0 (0.0) & 0.53 (0.14) &   1.5 (0.0) &   2.2 (0.4) \\ 
			&  & PMSc$_{\text{CV}}$-CoCo &  0.0 (0.0) &  0.0 (0.0) &  0.0 (0.0) &  0.0 (0.0) & 0.36 (0.13) &  23.4 (0.4) &   0.0 (0.0) \\ 
			&  & PMSc$_{\text{FS}}$-CoCo &  7.5 (0.0) &  0.0 (0.0) &  0.0 (0.0) &  0.0 (0.3) & 0.46 (0.11) &  5.4 (1.1) &  1.3 (0.1) \\ 
			&  & SISc-CoCo &  7.5 (0.0) &  0.0 (0.0) &  0.1 (0.0) &  0.0 (0.0) & 0.55 (0.12) &   1.5 (0.0) &   1.6 (0.2) \\ [3pt]
			& 2000 & One-stage Corrected &  - &  - &  1.2 (0.5) &  0.0 (0.0) & 0.61 (0.13) &  0.0 (0.0) & 45.7 (23.4) \\ 
			&  & PMSc$_{\text{CV}}$-Corrected &  0.0 (0.0) &  0.0 (0.0) &  0.0 (0.0) &  0.0 (0.0) & 0.37 (0.13) & 37.6 (1.6) &  0.3 (0.1) \\ 
			&  & PMSc$_{\text{FS}}$-Corrected &  3.8 (0.0) &  0.0 (0.0) &  0.5 (0.3) &  0.0 (0.0) & 0.45 (0.12) & 17.2 (3.0) &  2.4 (0.4) \\ 
			&  & SISc-Corrected &  3.8 (0.0) &  0.0 (0.0) &  0.5 (0.3) &  0.0 (0.0) & 0.55 (0.13) &  2.8 (0.1) &  1.9 (0.3) \\ 
			&  & PMSc$_{\text{CV}}$-CoCo &  0.0 (0.0) &  0.0 (0.0) &  0.0 (0.0) &  0.0 (0.9) & 0.37 (0.17) & 37.6 (1.6) &  0.0 (0.0) \\ 
			&  & PMSc$_{\text{FS}}$-CoCo & 3.8 (0.0) &  0.0 (0.0) &  0.1 (0.0) &  0.1 (1.3) & 0.46 (0.12) & 17.2 (3.0) &  1.7 (0.2) \\ 
			&  & SISc-CoCo &  3.8 (0.0) &  0.0 (0.0) &  0.1 (0.0) &  0.0 (0.0) & 0.56 (0.13) &  2.8 (0.1) &  1.4 (0.1) \\ 
			\bottomrule[1.5pt]
	\end{tabular}}
\end{table}


\begin{table}[ht]
	\centering
	\caption{Performance of the one-stage and two-stage estimators in the simulation study based on false positive rate (FPR, in percentage), false negative rate (FNR, in percentage), $\ell_2$ estimation error, and computation time (in seconds) when $\bm{\Sigma}_x$ has an homogeneous structure with $\rho_x = 0.3$. Standard error are included in parentheses. The CoCo estimator was only computed when the number of variables was no more than $1000$ (either in the 1st or 2nd step). }
	\medskip
	\resizebox{\linewidth}{!}{\begin{tabular}{lll rr rrr rr}
			\toprule[1.5pt]
			$\bm\Sigma_u$  & $p$ & Estimator &  \multicolumn{2}{c}{1st step} &
			\multicolumn{3}{c}{2nd step} & \multicolumn{2}{c}{Time}\\
			
			& & & FPR  & FNR  & FPR & FNR & $\ell_2$& 1st step & 2nd step \\ 
			\cmidrule(lr){4-5} \cmidrule(lr){6-8} \cmidrule(lr){9-10}	
			
			Diagonal & 1000& One-stage Corrected &   - &   - &  20.0 (29.0) &   0.1 (1.5) & 0.96 (0.77) &   0.0 (0.0) &  18.9 (3.3) \\ 
			&  & PMSc$_{\text{CV}}$-Corrected & 100.0 (0.0) &   0.0 (0.0) &  19.7 (28.8) &   0.2 (2.0) & 0.95 (0.77) &  12.7 (0.7) &  18.7 (3.3) \\ 
			&  & PMSc$_{\text{FS}}$-Corrected &  7.5 (0.0) &  1.2 (4.8) &  3.8 (3.4) &  1.2 (4.8) & 1.12 (0.85) &   6.0 (1.1) &   0.9 (0.1)   \\ 
			&  & SISc-Corrected &   7.6 (0.1) &   6.7 (10.5) &   1.0 (0.9) &   6.7 (10.5) & 0.54 (0.30) &   1.3 (0.1) &   1.0 (0.2) \\ 
			&  & One-stage CoCo &   - &   - &  11.5 (2.2) &   0.0 (0.0) & 0.77 (0.13) &   0.0 (0.0) & 481.5 (27.5) \\ 
			&  & PMSc$_{\text{CV}}$-CoCo & 100.0 (0.0) &   0.0 (0.0) &  11.5 (2.2) &   0.0 (0.0) & 0.77 (0.13) &  12.7 (0.7) & 476.7 (26.8) \\ 
			&  & PMSc$_{\text{FS}}$-CoCo &   7.5 (0.0) &  1.2 (4.8) &  0.8 (0.3) &  1.2 (4.8) & 0.36 (0.12) &   6.0 (1.1) &   0.9 (0.2) \\ 
			&  & SISc-CoCo &   7.6 (0.1) &   6.7 (10.5) &   1.0 (0.4) &   6.7 (10.5) & 0.52 (0.23) &   1.3 (0.1) &   1.6 (0.2) \\ [3pt]
			& 2000 & One-stage Corrected &   - &   - &  29.1 (21.0) &   1.1 (5.1) & 1.57 (0.84) &  0.0 (0.0) & 49.9 (7.0) \\ 
			&  & PMSc$_{\text{CV}}$-Corrected & 100.0 (0.0) &   0.0 (0.0) &  28.3 (21.3) &   1.0 (5.1) & 1.53 (0.84) & 25.2 (1.4) & 49.3 (6.9) \\ 
			&  & PMSc$_{\text{FS}}$-Corrected &   3.8 (0.0) &   1.1 (4.5) &   1.8 (1.7) &   1.1 (4.5) & 1.09 (0.86) & 10.1 (1.7) &  0.7 (0.1) \\ 
			&  & SISc-Corrected &   3.8 (0.0) &  11.0 (13.1) &   0.5 (0.4) &  11.0 (13.1) & 0.61 (0.31) &  2.7 (0.1) &  0.9 (0.2) \\ 
			&  & PMSc$_{\text{CV}}$-CoCo & 100.0 (0.0) &   0.0 (0.0) &   - &   - &  - & 25.2 (1.4) &  - \\ 
			&  & PMSc$_{\text{FS}}$-CoCo &   3.8 (0.0) &   1.1 (4.5) &   0.4 (0.2) &   1.1 (4.5) & 0.33 (0.11) & 10.1 (1.7) &  0.9 (0.2) \\ 
			&  & SISc-CoCo &   3.8 (0.0) &  11.0 (13.1) &   0.5 (0.3) &  11.0 (13.1) & 0.60 (0.29) &  2.7 (0.1) &  1.9 (0.2) \\ \\
			
			\multirow{2}{1.2cm}{Block diagonal} & 1000 & One-stage Corrected &   - &   - &  23.5 (29.0) &   1.1 (5.2) & 1.25 (0.73) &   0.0 (0.0) &  22.7 (4.5) \\ 
			&  & PMSc$_{\text{CV}}$-Corrected & 100.0 (0.0) &   0.0 (0.0) &  25.6 (29.4) &   0.8 (4.2) & 1.30 (0.74) &  14.5 (0.7) &  22.5 (4.6) \\ 
			&  & PMSc$_{\text{FS}}$-Corrected &   7.5 (0.0) &   0.9 (4.4) &   5.3 (3.2) &   0.9 (4.4) & 1.56 (0.74) &   5.9 (1.1) &   1.3 (0.4) \\ 
			&  & SISc-Corrected &   7.6 (0.1) &  14.2 (14.8) &   1.5 (2.1) &  14.2 (14.9) & 0.97 (0.48) &   1.3 (0.1) &   1.1 (0.3) \\ 
			&  & One-stage CoCo &   - &   - &  14.3 (2.3) &   0.4 (3.2) & 1.23 (0.17) &   0.0 (0.0) & 417.2 (28.2) \\ 
			&  & PMSc$_{\text{CV}}$-CoCo & 100.0 (0.0) &   0.0 (0.0) &  14.3 (2.2) &   0.5 (3.3) & 1.23 (0.17) &  14.5 (0.7) & 413.4 (28.5) \\ 
			&  & PMSc$_{\text{FS}}$-CoCo &  7.5 (0.0) &   0.9 (4.4) &   0.7 (0.4) &   0.9 (4.4) & 0.57 (0.15) &   5.9 (1.1) &   2.2 (0.2) \\ 
			&  & SISc-CoCo &   7.6 (0.1) &  14.2 (14.8) &   0.9 (0.4) &  14.2 (14.8) & 0.84 (0.31) &   1.3 (0.1) &   2.3 (0.2) \\[3pt] 
			
			& 2000 & One-stage Corrected &   - &   - &  32.4 (17.1) &   3.0 (10.1) & 1.86 (0.65) &  0.0 (0.0) & 52.5 (9.7) \\ 
			&  & PMSc$_{\text{CV}}$-Corrected & 100.0 (0.0) &   0.0 (0.0) &  31.8 (17.4) &   3.3 (10.1) & 1.84 (0.65) & 27.7 (1.5) & 52.1 (9.6) \\ 
			&  & PMSc$_{\text{FS}}$-Corrected &   3.8 (0.0) &   2.0 (6.3) &   2.8 (1.5) &   2.0 (6.3) & 1.63 (0.71) & 10.1 (1.7) &  0.9 (0.3) \\ 
			&  & SISc-Corrected &   3.8 (0.0) &  21.4 (16.6) &   0.8 (1.0) &  21.4 (16.6) & 1.09 (0.48) &  2.8 (0.1) &  1.0 (0.3) \\ 
			&  & One-stage CoCo &   - &   - &   - &   - &  - &  0.0 (0.0) &  - \\ 
			&  & PMSc$_{\text{CV}}$-CoCo & 100.0 (0.0) &   0.0 (0.0) &   - &   - &  - & 27.7 (1.5) &  - \\ 
			&  & PMSc$_{\text{FS}}$-CoCo &   3.8 (0.0) &   2.0 (6.3) &   0.4 (0.2) &   2.0 (6.3) & 0.59 (0.16) & 10.1 (1.7) &  1.9 (0.2) \\ 
			&  & SISc-CoCo &   3.8 (0.0) &  21.4 (16.6) &   0.5 (0.3) &  21.4 (16.6) & 0.97 (0.36) &  2.8 (0.1) &  2.5 (0.2) \\ \bottomrule[1.5pt]
	\end{tabular}}
\end{table}

\label{lastpage}

\end{document}